\documentclass[%
 reprint,
 amsmath,amssymb,
 aps,
]{revtex4-2}
\usepackage{xcolor}
\newcommand{\todo}[1]{\textcolor{red}{\textbf{TODO: }} \emph{\color{red}#1}}

\usepackage{graphicx}
\usepackage{dcolumn}
\usepackage{bm}
\usepackage{siunitx}
\usepackage{listings} 
\newcommand{\qopt}[0]{\texttt{qopt}}

\definecolor{codegreen}{rgb}{0,0.6,0}
\definecolor{numbercolour}{rgb}{0.5,0.5,0.5}
\definecolor{stringcolour}{rgb}{0.,0.6,0.}
\definecolor{backcolour}{rgb}{0.95,0.95,0.95}

\lstdefinestyle{mystyle}{
    backgroundcolor=\color{backcolour},   
    keywordstyle=\color{blue},
    numberstyle=\tiny\color{numbercolour},
    stringstyle=\color{stringcolour},
    basicstyle=\ttfamily\footnotesize,
    breakatwhitespace=false,         
    breaklines=true,                 
    captionpos=b,                    
    keepspaces=true,                 
    numbers=left,                    
    numbersep=5pt,                  
    showspaces=false,                
    showstringspaces=false,
    showtabs=false,                  
    tabsize=2,
    language=Python,
}

\def\ContinueLineNumber{\lstset{firstnumber=last}}

\begin{document}

\preprint{APS/123-QED}

\title{\qopt{}: An experiment-oriented Qubit Simulation and Quantum Optimal Control Package}

\author{Julian D. Teske}
\email{j.teske@fz-juelich.de}
 \affiliation{JARA-FIT Institute for Quantum Information, Forschungszentrum J\"ulich GmbH and RWTH Aachen University, 52074 Aachen, Germany}

\author{Pascal Cerfontaine}
\affiliation{JARA-FIT Institute for Quantum Information, Forschungszentrum J\"ulich GmbH and RWTH Aachen University, 52074 Aachen, Germany}%

\author{Hendrik Bluhm}
\email{bluhm@physik.rwth-aachen.de}
\affiliation{JARA-FIT Institute for Quantum Information, Forschungszentrum J\"ulich GmbH and RWTH Aachen University, 52074 Aachen, Germany}%

\date{\today}

\begin{abstract}
Realistic modeling of qubit systems including noise and constraints imposed by control hardware is required for performance prediction and control optimization of quantum processors.
We introduce \qopt{}, a software framework for simulating qubit dynamics and robust quantum optimal control considering  common experimental situations. To this end, we model open and closed qubit systems with a focus on the simulation of realistic noise characteristics and experimental constraints. Specifically, the influence of noise can be calculated using Monte Carlo methods, effective master equations or with the efficient filter function formalism, which enables the investigation and mitigation of auto-correlated noise. In addition, limitations of control electronics including finite bandwidth effects as well as nonlinear transfer functions and drive-dependent noise can be considered. The calculation of gradients based on analytic results is implemented to facilitate the efficient optimization of control pulses. The software easily interfaces with QuTip, is published under an open source license, well-tested and features a detailed documentation.
\end{abstract}

\maketitle


\section{Introduction}

A central challenge for the realization of a universal quantum computer is the loss of entanglement and coherence due to the detrimental effects of uncontrolled interactions between the system storing quantum information and its environment \cite{Unruh95}. 
To understand the underlying error processes and predict the performance of quantum processor designs, a realistic model for the manipulation by control hardware and the noise introduced by the environment is required. Such a model is also essential for optimizing the performance, for example using  quantum optimal control (QOC) techniques,  which adapt control pulses such that a suitably chosen metric is maximized \cite{TrainSchroedCat, qoc_outlook_2010}. 
QOC methods will be essential to achieve the best possible gate accuracy for quantum devices, 
both for noisy intermediate-scale systems and future universal quantum computers \cite{optimizedPulsesTransmon, OQCwithrandomizedbenchmarking, Preskill2018, Wang2014, Yang2016, Anderson2015}. 
For increasing the number of qubits in a quantum processing unit, QOC techniques will likely be needed to address effects like crosstalk of control fields, frequency crowding and unintended coupling between qubits. Furthermore, a well-founded performance assessment for benchmarking, platform selection and system design is only possible if the best possible control approaches are considered. 

In a physical system, the time-dependent control fields implementing a  desired quantum gate, which we refer to as control pulses, are applied to a qubit using classical control hardware (such as arbitrary waveform generators or lasers). 
In numerical QOC, a qubit's evolution is simulated for a given control pulse and then optimized prior to the application to the experiment.
To achieve the best possible qubit performance, this optimization requires a model of the whole system including any effects  that influence  its evolution. These factors can include the intrinsic properties of the qubits themselves, their coupling, the unintended interaction with the environment  \cite{Floether_2012MarkVsNonMark} as well as their response to and properties of the control hardware. 
Incorporating the control hardware into the models is also essential for the design of tailored control electronics for quantum computation, e.g, cryolectronic systems operated in close vicinity of the qubits for achieving a high wiring density. While providing at least the minimally required control capabilities, additional constraints like a reduced heat dissipation compatible with cryogenic cooling need to be considered \cite{Geck_2019, CharbonCryoCMOS} for such systems.

Even if the models used for pulse optimization are not sufficiently accurate for the direct ("open loop") experimental application of the results, they can be a useful starting point for further fine-tuning in a closed loop with feedback from an actual experiment \cite{PascalNatureComms, PascalGSCTheo, WilhelmAdHOC2014}. Even then, they should behave as similar as possible to the physical system and at least qualitatively capture all relevant effects. Simulation frameworks are also helpful to simulate and develop such experimental tuning procedures \cite{PascalHighFidGates, PascalHighFidSingleQubitGatesTheo, WilhelmAdHOC2014}, or they can be combined with the characterization of a qubit in an optimization loop \cite{C3wittler2020integrated}.

The desire to achieve realistic models leads to a number of required features.
The signal transduction from the hardware to the qubit must be modeled including all 
relevant technological limitations, such as the finite bandwidth of arbitrary waveform generators \cite{TransferFunc} and signal pathways, that effect the actually implemented gate, can also cause gate-bleedthrough \cite{KellyBT2014} and crosstalk between different control signals.
Furthermore, nonlinear effects may arise, either due to the control-hardware itself, or because of a nonlinear relation between the physical control fields and the effective control Hamiltonian, e.g., due to a truncation of the Hilbert space involving a Schrieffer-Wolf transformation. 
Regarding decoherence effects, it is important to realistically consider all relevant noise sources and their properties. An important aspect are correlations of noise, which can be incorporated by nontrivial spectral noise densities \cite{Floether_2012MarkVsNonMark}.  Correlated noise is a limiting factor for high-fidelity gates in many systems  \cite{ChargeNoiseSpectrPhysRevLett.110.146804, PascalHighFidGates, 999Yoneda2018}, but correlations can also be exploited to reduce decoherence effects in dynamically corrected gates.
The performance quantification of quantum operations can be extended to such a mitigation of noise effects, often termed robust optimal control \cite{Floether_2012MarkVsNonMark}. 
In case of nonlinear relations between control fields and qubit Hamiltonian, the effect of noise in the control field automatically depends on the control field, thus representing drive-dependent noise.


From a performance point of view, it is important to be able to use efficient algorithms that are well-suited for the problem at hand.
Many algorithms for the numerical optimization discretize control pulses in time to yield a finite dimensional parameter space. The discrete elements of the control pulses can then be updated simultaneously using gradient ascent methods such as GRAPE \cite{grape} or subsequently with Krotov's method \cite{Krotov_Schirmer_2011} as well as gradient-free methods \cite{Huang2017}. More advanced gradient-based algorithms include the use of second order derivatives \cite{KuprovSecondOrderGradient}, gradient optimization of analytical controls (GOAT) \cite{goat} and the application of the Kalman filter for the estimation of gradients \cite{TeskeKalman}. Another approach is to parameterize control pulses in terms of a randomly chosen subspace using CRAB \cite{crab} or the remote version RedCRAB \cite{RedCRABHeck}. 

To ease the application of advanced models and algorithms, general purpose, flexible and easily usable software implementations are highly advantageous. 
An early example is the unifying algorithmic framework DYNAMO \cite{DYNAMO}, which implements GRAPE and Krotov's method in Matlab and inspired the implementation of an optimal control package in the Quantum Toolbox in Python (QuTiP) \cite{QUTIP}, an open-source software for the simulation of open quantum systems. An additional package introduces Krotov's method to QuTiP as described by Goerz \textit{et al.} \cite{Krotov}. QuTiP's extension by the subpackage for quantum information processing introduces the capability to simulate quantum gates on the pulse level with the option to include noise but without optimal control techniques. There are also special purpose optimization frameworks like QEngine \cite{SORENSEN2019135QEngine} for ultracold atoms or Pulser \cite{pulser} for neutral-atom arrays.
Some of these implementations can be generalized to noisy systems if an open system description based on master equations is adopted \cite{Koch_2016robustopensys}, thus readily treating Markovian noise.
One possibility do deal with non-Markovian noise is the use of ancillary qubits \cite{Floether_2012MarkVsNonMark, Pawela2016methodsdecoherence}, which however is computationally very costly as it substantially increases the Hilbert space dimension. A methodology combining open and closed loop optimization is the C3 tool set for integrated control, calibration and characterization \cite{C3wittler2020integrated}.

While these simulation frameworks are widely and successfully used in their respective domains, we found them to be less suited and difficult to extend to address the above requirements for the realistic, hardware aware simulation. We thus implemented the new python package \qopt{} \cite{qopt}, which was in many ways inspired by QuTiP's optimal control subpackage, but has in some aspects a different structure.
Specifically, the performance of sequential optimization algorithms like Krotov's method is based on the possibility to efficiently update pulses independently in each time step, which is incompatible with our current implementation of parameterized pulses and transfer functions.


Concurrent with our work, the startup Q-CTRL developed a software with similar methods but pursued a different strategy and targeted a commercial audience \cite{QCTRLball2020software}. As for \qopt{}, these methods include generalized filter functions and the simulation of noise by explicitly sampling noise distributions as Monte Carlo simulations. The imperfections of control hardware are modeled by transfer functions. Additionally, Q-CTRL provides methods for the noise characterization of a given system.
While one may expect that the commercial multi-purpose software of Q-CTRL leads to a feature-rich and easy to use solution, the closed-source approach reduces the transparency and flexibility, which is often important for research use.
As an GPL3-licensed open-source package, \qopt{} complements this approach, targeting mainly the scientific community in academia and industry. Besides low barriers to entry, the modular structure and complete API documentation \cite{qopt-docs} provide full flexibility in the implementation of new techniques that expand the application to unsolved problems. The user can supply her own optimization algorithm or replace any other relevant part. Multi-processing is also supported.

This paper gives an overview of \qopt{}'s capabilities while a full documentation and numerous introductory examples can be found online on readthedocs \cite{qopt-docs}. 
Section \ref{sec:problem and methods} describes the mathematical formulation used for the experimentally oriented simulation of qubits and the application in QOC. The actual implementation of the \qopt{} package is portrayed in Section \ref{sec::implementation}, including a practical example. Finally, an outlook is given in Section \ref{sec:summary and outlook}.

\section{Problem formulation and simulation methods}
\label{sec:problem and methods}

\subsection{Problem formulation}
A rather general model for a driven qubit system subject to (classical) noise can be described by a Hamiltonian of the form
\begin{align}
    H(t) &= H_c(t) + H_d(t) + H_n(t) \\
    H_c(t) &= \sum_k u_k(t) C_k \label{eq::hamiltonian} \\
    H_n(t) &= \sum_k b_k(t) s_k(t) C_k
\end{align}
with the control Hamiltonian $H_c$, the drift Hamiltonian $H_d$ and the noise Hamiltonian $H_n$. 
The control Hamiltonian models the manipulation of the system with time dependent control amplitudes $u_k$ and Hermitian operators $C_k$. 
The drift Hamiltonian $H_d$ incorporates any effects that cannot be freely controlled but still affect the dynamics. It describes the natural evolution of the system in absence of any control, e.g., due to a fixed energy splitting.
The noise Hamiltonian $H_n$ models unintentional interactions of the system with the control hardware or the environment, like electrical noise on the control amplitudes or the interaction with electromagnetic fields from the host material. The noise amplitudes $b_k$ describe the strength of the perturbation while the noise susceptibilities $s_k$ describe the coupling strength to the noise source. $s_k$ can depend on the control amplitudes $u_k$ to model noise originating from the control mechanism. We assume the noise to be classical with zero mean and (wide-sense) stationary. Auto-correlated classical noise is characterized (up to higher-order correlations for non-Gaussian processes) by its spectral density $S_k(\omega)$ defined as the Fourier transform of the correlator $\langle b_k(t_1)b_k(t_2) \rangle = \frac{1}{2 \pi} \int_{- \infty}^{\infty} S_k(\omega) e^{-i\omega (t_1 - t_2)}  d\omega$.


In many experiments the control amplitudes $u_k(t)$ in the Hamiltonian from Eq.~\eqref{eq::hamiltonian} are not directly controllable by the experimentalist. Rather, they are functions of physical control fields $v_i(t)$. We call the mapping of $v_i$ to $u_k$ the \emph{amplitude function} (see Fig. \ref{fig:activity diagram}). An example is the control by Rabi-driving where the control amplitude $v_1(t) = A(t)$ and phase offset $v_2(t) = \delta(t)$ of a control signal appear in the Hamiltonian as $u_1(t) = A(t) \text{sin}(\omega t + \delta(t))$. Nonlinear amplitude functions can also arise from the truncation of Hilbert space. For example, the exchange interaction of two electron spins depends nonlinearly on the detuning between different orbital states of the electrons, when the orbital states are truncated.

Furthermore, imperfections of the control electronics can be modeled by the use of linear transfer functions \cite{TransferFunc} acting on the controllable optimization parameters $v_k$. This can be done by oversampling and smoothing the control pulse e.g., by convolution with a Gaussian kernel or by using a realistic transfer function, measured for a specific device. Our implementation also allows the user to add boundary conditions by padding the beginning and end of each pulse with appropriate values. A common use case is an additional idle time at the end of each pulse in order to avoid transients across pulses \cite{PascalHighFidGates}.
Note that the amplitude and transfer functions have similar roles, but different constraints. The amplitude function can be nonlinear but must be local in time, whereas the transfer function must be linear but can be nonlocal in time. The transfer function is applied before the amplitude function.

\subsection{Noise Simulation}
For the numerical solution of the Schroedinger equation, we assume piecewise constant control during $n_t$ time steps of length $(\Delta t_1, \dots, \Delta t_{n_t})$. The total unitary propagator of an evolution is calculated as a product of matrix exponentials of the time-independent Hamiltonians $U = e^{-i H(t_{n_t})\Delta t_{n_t}} \dots e^{-i H(t_2)\Delta t_2} \cdot e^{-i H(t_1)\Delta t_1}$ using the convention $\hbar = 1$.

Noise can be taken into account with several methods, which might be more or less appropriate and numerically efficient depending on the noise properties.

Explicitly generating numerous noise traces whose Fourier transform converges (on average) to the noise spectral density $S$ is one of the simplest methods. In this approach the highest relevant noise frequency sets the required time step for the numerical integration, so that additional oversampling is required if it exceeds the bandwidth of the pulse itself. 
Since the numerical complexity of the simulation grows proportionally with oversampling, such a Monte Carlo approach becomes computationally inefficient if the noise spectral density cannot be neglected at frequencies much higher than the bandwidth frequency of the control electronics. Even if this is not an issue, many repetitions are required to gather statistics.

If the main noise contribution occurs at frequencies much lower than the simulation dynamics, it is sufficient to consider the noise amplitude to be static during the pulse. For few noise sources, explicit numerical integration of the (typically Gaussian) probability distribution of these noise values is more efficient than Monte Carlo sampling in small dimensions \cite{CaflischMonteCarloGeneral}. The user can choose between both methods in \qopt{}.

If high noise frequencies are relevant, more efficient methods than Monte Carlo or numerical integration are available. They are based on  master equations and filter functions for white and auto-correlated noise described by spectral densities, respectively \cite{ConvertLindblad, Green_2013, PascalHighFidSingleQubitGatesTheo, PascalHighFidGates}. 
In the special case of Markovian (uncorrelated) noise, the influence on the qubit system can be described by an effective master equation in Lindblad form \cite{ConvertLindblad}. In this master equation, the von-Neumann equation is complemented by a dissipation term leading to non-unitary dynamics. The Lindblad form is
\begin{align}
    \dot{\rho} = -\frac{i}{\hbar}\left[ H, \rho \right] + \sum_n \gamma_n \left(L_n \rho L_n^\dagger - \frac{1}{2}\{ L_n^\dagger L_n, \rho \}\right),
\end{align}
where the Lindblad operators $L_n$ describe dissipation effects and can themselves depend on the control amplitudes.

The master equation is written as linear system of equations
with the Kronecker matrix product $\otimes$ and calculated as matrix exponential:
\begin{align}
\text{vec}(\rho) (t) & = \exp[(-i\mathcal{H} + \mathcal{G})t] \; \text{vec}(\rho)(0), \\
\mathcal{H} &=  I \otimes H - H^T \otimes I, \\
\mathcal{G} &= \sum^K_{k=0} \mathcal{D}(L_k), \\
\mathcal{D}(L) &= L^\ast \otimes L - \frac{1}{2} I \otimes (L^\dag L) - \frac{1}{2} (L^TL^\ast) \otimes I,
\end{align}
where $\text{vec}(\rho)$ denotes the density matrix written as vector in column-wise order. 

The derivation of a master equation in Lindblad form requires the assumption of Markovian (uncorrelated) noise. This approximation does not hold for many experimentally relevant noise sources such as flux noise in superconducting qubits and charge noise in many types of solid state qubits, which typically have an $1/f$-like spectrum. 
The filter function formalism provides a mathematical tool which can (perturbatively) model the decoherence caused by arbitrary classical noise \cite{Green_2013}. Both master equation and filter functions have already been employed for numerical pulse optimization \cite{ffoptPhysRevA.99.042310, PascalHighFidSingleQubitGatesTheo}.

In the filter function formalism, a so-called filter function $F_\alpha$ can be calculated given the evolution of the system for each noise source $\alpha$. For a given control pulse, $F_\alpha$ captures the susceptibility of the resulting quantum channel to the noisy quantity as function of frequency.
The noise contribution to the entanglement infidelity or other figures of merit can be calculated as the integral of the filter function and the spectral noise density
\begin{align} \label{eq:entanginfidff}
    \mathcal{I}_{\text{ff}} = \int_{-\infty}^{\infty} \frac{d\omega}{2 \pi} S_\alpha(\omega) F_\alpha(\omega).
\end{align}

The filter function formalism is an approximate and efficient method to calculate the infidelity caused by fast non-Markovian noise of small amplitude. It outperforms Monte Carlo simulations for small systems, while Monte Carlo methods scale better with an increasing number of qubits \cite{hangleiter2021filter, cerfontaine2021filter}. The numerical routines for the calculation of filter functions and their derivatives with respect to the control amplitudes $\frac{\partial F_\alpha}{\partial u_k(t)}$ are provided by the software package filter\_functions \cite{filter_func_package}.

\subsection{Fidelity measures}
\qopt{} implements various fidelity measures to quantify the accuracy of quantum operations. For state transfers, the state fidelity between the initial and final quantum states described by the density matrices $\rho_1$ and $\rho_2$ can be used, which is defined as
$
    \mathcal{F}_{\text{st}} (\rho_1, \rho_2) = \left[\text{tr}\left (\sqrt{\sqrt{\rho_1}\rho_2 \sqrt{\rho_1}}\right)\right]^2.
$

One commonly used measure for the closeness of two quantum gates is the entanglement infidelity $\mathcal{I}_{\text{e}}(V^\dag \circ U)$. If both the simulated propagator $U$ and the target gate $V$ are unitary, the entanglement fidelity $\mathcal{F}_{\text{e}} = 1- \mathcal{I}_{\text{e}}$ is given by the Hilbert-Schmidt norm as
\begin{align}
    \mathcal{F}_{\text{e}}(V^\dag \circ U) = \frac{1}{d^2} |\text{tr}(V^\dag U)|^2.
\end{align}
This fidelity can also be generalized for open quantum systems \cite{Grace_2010openfidelity} and we calculate it as
\begin{align}
    \mathcal{F}_o(V, U_s) &= \frac{1}{d^2}  \text{tr}\left( \left(V^T \otimes V^\dag \right) U_s \right),
\end{align}
where $U_s$ is the simulated noisy quantum process. 

Leakage occurs in a quantum processor if states outside the computational subspace are populated.
To simulate this error source, we extend the Hilbert space of computational states $\mathcal{H}_c$ as vector space sum $\mathcal{H} = \mathcal{H}_c + \mathcal{H}_l$ by the space $\mathcal{H}_l$ spanned by the leakage states. We quantify the amount of information lost into the leakage subspace by cropping the unitary evolution on the entire system to the computational states and calculating the distance to unitarity of the projected propagator $U_c$ as
\begin{align}
    \mathcal{L} = 1 - \text{tr}(U_c^\dag U_c) / d.
\end{align}
In order to investigate the amount of incoherent leakage, i.e., caused by noise, this cost function can be combined with a noise simulation.

\subsection{Optimization Procedures}

The minimization of such a fidelity over the optimization space spanned by all possible control pulses $u(t)$ formally defines QOC as the minimization problem:
\begin{align}
    \min_{u(t)} \mathcal{I}(u(t)) = \mathcal{I} (u^\ast(t)).
\end{align}

Although \qopt{} provides a general interface for cost functions that can be used with any optimization algorithm, we set a special focus on the use of gradient-based optimization algorithms. They are widely used and also applicable in QOC \cite{grape}. For this purpose we implement the analytic calculation of exact gradients, which do not require any assumption about the time discretization nor about the control strength and represent an alternative to the use of automatic differentiation \cite{C3wittler2020integrated, TrajectoriesAutodiff}.

When multiple cost functions are evaluated, the software supplies them as a vector to the optimization algorithm. This leaves more options for the optimization and allows to give each cost function an individual weight.


\section{Implementation}
\label{sec::implementation}

\begin{figure*}
    \centering
    \includegraphics[width=\textwidth]{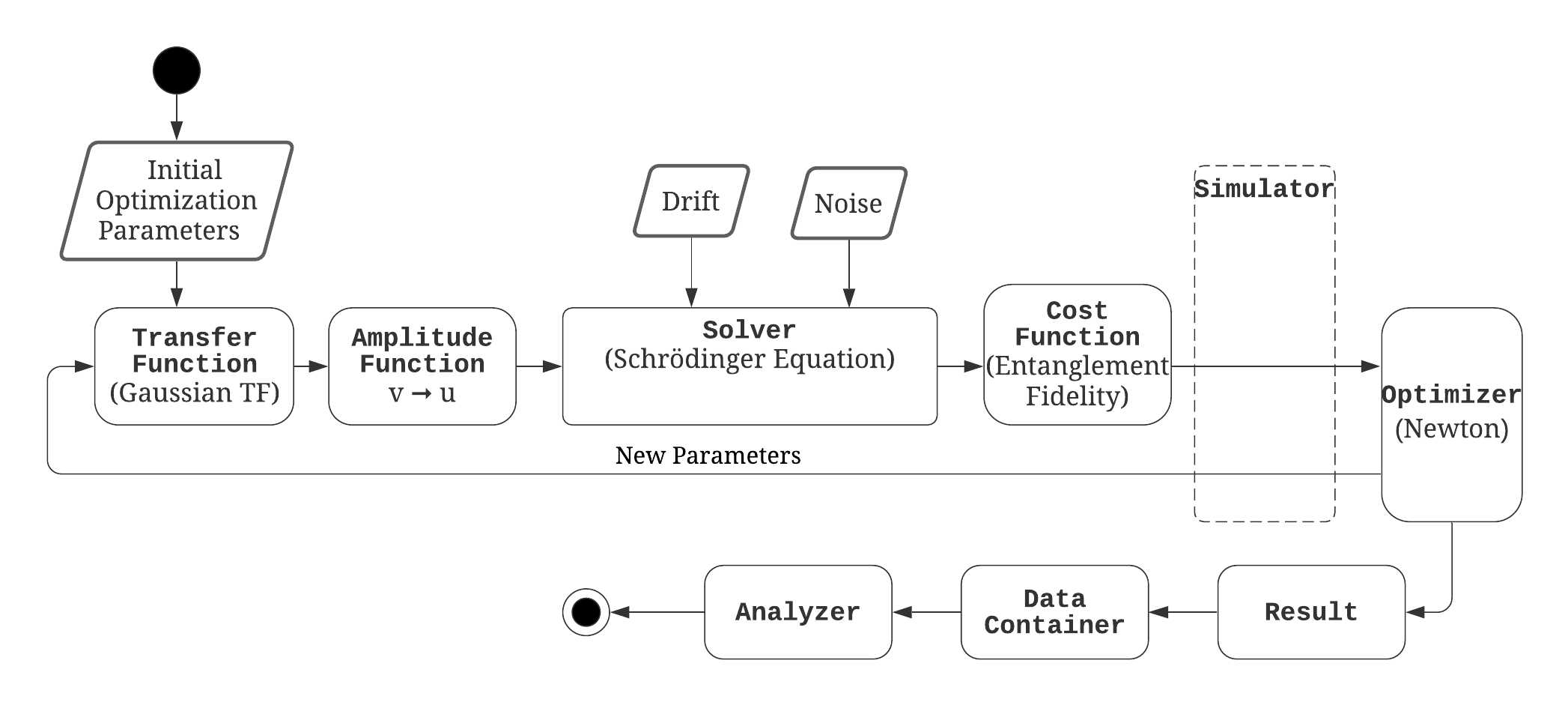}
    \caption{Activity diagram of a pulse optimization with \qopt{}. The solid and encircled dots show the start and end point respectively. The arrows mark the flow of activity and the rounded rectangles the classes of the \qopt{} package with base class names written in bold and examples in brackets, while the parallelograms contain direct input from the model. The box of the simulator is dashed because it defines an interface between the simulation and the optimization. Concrete examples or explanations are given in the brackets below the class name, i.e., the example for a transfer function is the convolution with a Gaussian kernel.}
    \label{fig:activity diagram}
\end{figure*} 

In this section we present \qopt{}'s object-oriented implementation by first discussing the structure of an optimal control setup with \qopt{}, and then outlining the optimization of an $X_{\pi / 2}$ gate for a single qubit controlled by Rabi-driving as a simple illustrative example.

\subsection{Program Flow}

The intended setup is plotted as an activity diagram in Fig.~\ref{fig:activity diagram} giving a full picture, although not every feature and every class needs to be used in practice. Only the solver class is central to the simulation. The diagram shows the modular software structure and the flow of information between the classes.

The optimization commences with a set of initial optimization parameters $v_k$, which can be chosen randomly or based on some insight. The package also features convenience functions to run the simulation with many different initial conditions in parallel to exploit the problem structure for trivial full parallelization. 

First, the ideal pulse parameters are mapped to the actual pulse seen by the qubits as defined by a transfer function class. Then, the control amplitudes $u_k$ are calculated by the amplitude function class.
The control amplitudes enter the Schroedinger or master equation in Lindlbad form together with the drift dynamics and the noise. The appropriate differential equation to describe the system is chosen by selecting the solver algorithm class. The solution of the differential equation is subsequently passed to the cost function class, to calculate the figure of merit for the optimization.

The simulator class is encircled in Fig.~\ref{fig:activity diagram} by a dashed box because it provides the interface between this simulation and the optimization algorithm. Furthermore it gathers run time statistics like the time spent in each cost function and for the calculation of the gradient. The optimizer class uses this interface to run the simulation in a loop until the internal termination criteria are met. Then it saves the final state in a result class and passes it to the data container class. The analyzer class can be used to visualize the results of several optimization runs stored in the data container class.

The object-oriented modular implementation of the code allows the user to easily replace single parts of the optimization framework. Among other things, this allows the user to make changes to the cost function (i.e., to use a different fidelity metric) or use a specific transfer or amplitude function. With the interface provided by the simulation class it is possible to use most standard optimization algorithms. Currently, the 'minimize' and 'least squares' functions of scipy's optimization subpackage are supported \cite{scipy}.

Numeric operations are encapsulated in an operator class. The computationally most expensive single operation during the simulation is the calculation of a matrix exponential as required for the numeric solution of the differential equations. The encapsulation allows the exchange of algorithms for the calculation of this matrix exponential (see for examples \cite{19matrixexponentials}). The Qobj class from QuTiP can be converted automatically into the \qopt{} operator class to improve the compatibility between both packages. This makes it easy to transfer simulations to \qopt{}.

More information about \qopt{} can be found in the online documentation \cite{qopt-docs}. It features a complete API reference documentation and two series of IPython notebooks. The former explains each feature of \qopt{} in detail while the latter discusses practical examples including some information about the physics and numerics of qubit simulation, how noise sources can be characterized, which type of noise simulation is the most efficient in each case and which effects noise will have on the system. These notebooks also serve as integration tests by demonstrating the consistency of different methods and the comparison with analytic calculations. Together with various unit tests for the critical parts of the implementation, they ensure the reliability of \qopt{}.


\subsection{Example}

\begin{figure}
    \centering
    \includegraphics{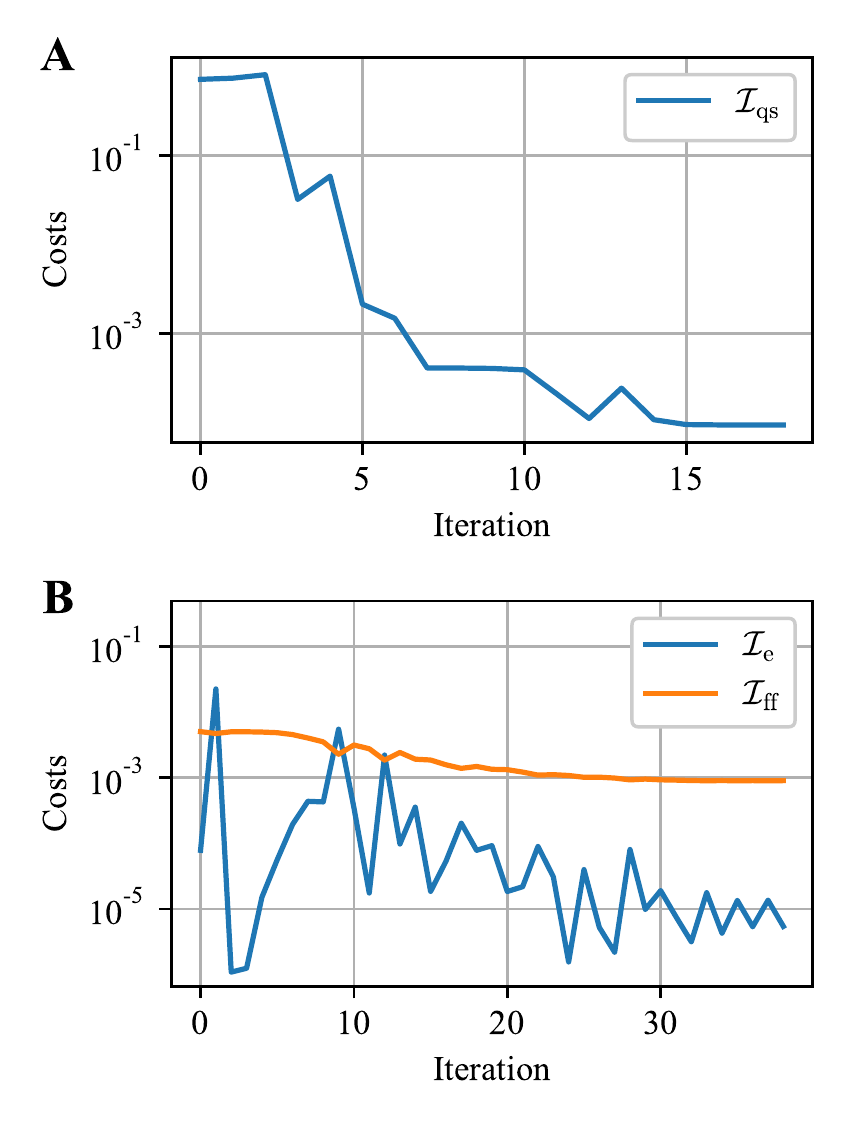}
    \caption{Infidelities as cost functions of the iteration during the optimization. (A) During the first optimization, we minimize the infidelity in a Monte Carlo simulation of quasi-static noise $\mathcal{I}_{\text{qs}}$ to find a pulse which is not susceptible to slow noise. (B) Subsequently, we use the final parameters of the first optimization to optimize the pulse for pink noise. We can see that the infidelity $\mathcal{I}_{\text{ff}}$ of Eq.~\eqref{eq:entanginfidff} decreases during the second optimization by about a factor of six. Thus pulses that mitigate slow noise are far from perfect in mitigating pink noise.}
    \label{fig:optimization_example}
\end{figure}
\begin{figure}
    \centering
    \includegraphics{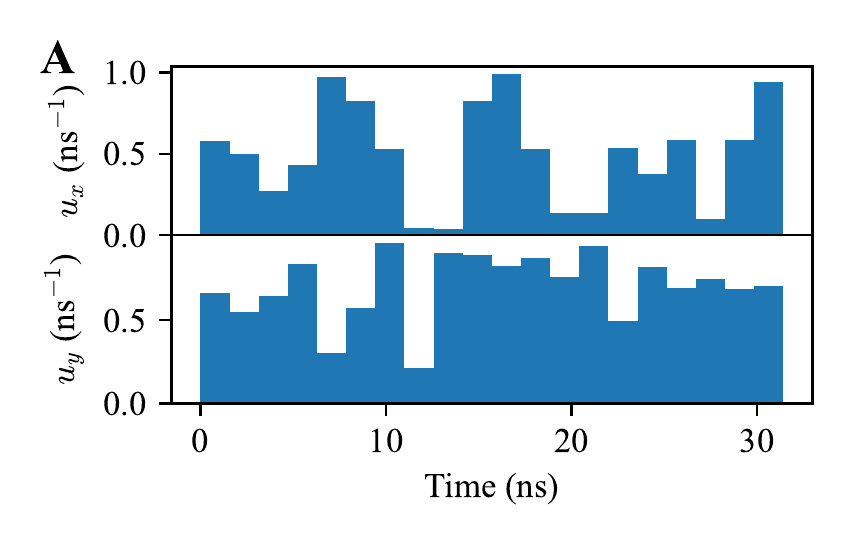}\\
    \includegraphics{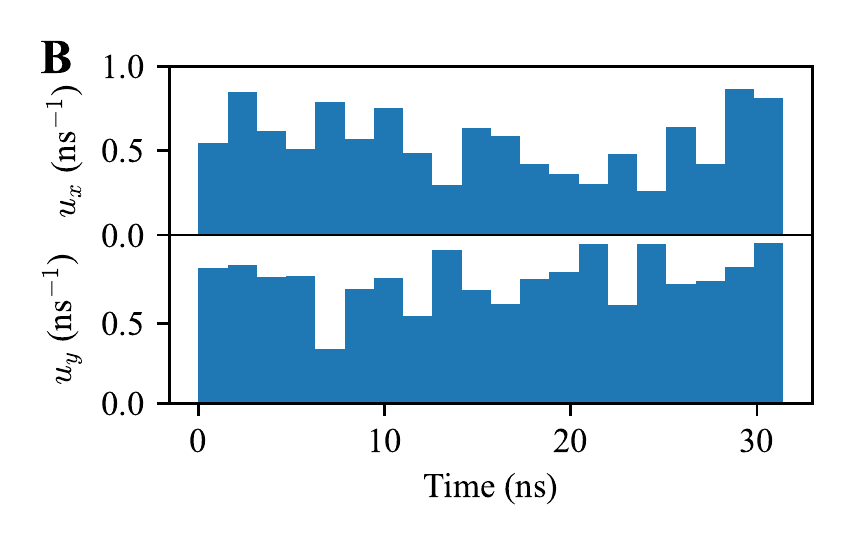}
    \caption{Plots of the optimized control amplitudes as function of time. (A) In the first optimization run, the control pulse is optimized for quasi-static noise. (B) The control amplitudes of the first run are used as starting point for an optimization in the presence of pink noise.}
    \label{fig:pulse_plots}
\end{figure}

We illustrate the usage of \qopt{} in the pulse optimization with the example of an $X_{\pi/2}$ single-qubit gate and optimize the pulse separately for quasi-static and auto-correlated fast noise, so that we can demonstrate how different noise models are implemented in \qopt{}'s API. 

The Hamiltonian of the example will be a single qubit manipulated by Rabi-driving in the rotating frame, which can be denoted as $H = u_x(t) \sigma_x + u_y(t) \sigma_y + \delta_\omega(t) \sigma_z$, where $u_x(t)$ and $u_y(t)$ are the control amplitudes (corresponding to the in-phase amplitude I and the quadrature amplitude Q of quadrature amplitude control), $\delta_\omega(t)$ the deviation of the driving frequency from the resonance frequency and the Pauli matrix $\sigma_i$ for $i \in {x,y,z}$. 

The detuning $\delta_\omega(t)$ enters Schrodinger's equation as a stochastic variable. In a first optimization we assume that the resonance frequency changes much slower than our gate time and is thus assumed to be constant during the pulse ($\delta_\omega(t) =\delta_\omega$). We therefore integrate $\delta_\omega$ over a Gaussian distribution and calculate the corresponding infidelity $\mathcal{I}_{\text{qs}}$ using a Monte Carlo method. 

In the second optimization we will use the final parameters of the first optimization as the initial pulse and assume that the resonance frequency is subject to pink noise and therefore the power spectral density of $\delta_\omega(t)$ has the form $S(f) = S_0 / f$. In this case we calculate the entanglement infidelity caused by systematic deviations $\mathcal{I}_{\text{e}}$ separately from the one caused by noise, which we calculate with filter functions $\mathcal{I}_{\text{ff}}$ as in Eq.~\eqref{eq:entanginfidff}.

\ContinueLineNumber

We optimize a pulse with 20 time steps of equal length:
\begin{lstlisting}
import qopt as qo
import numpy as np
import matplotlib.pyplot as plt

n_time_steps = 20
delta_t = .5 * np.pi
\end{lstlisting}

We start setting up the first simulation with quasi-static noise. The noise trace generator (NTG) provides the noise samples for the solver algorithm:

\begin{lstlisting}
noise_gen = qo.NTGQuasiStatic(
    n_samples_per_trace=n_time_steps,
    n_traces=10,
    standard_deviation=[.05, ]
)
\end{lstlisting}

The solver class holds the information about the Hamiltonian including the corresponding noise trace generator. The quantum operators are represented by the dense operator class that is based on a numpy array. We can also choose the exponential method, which is the algorithm used to calculate the matrix exponential and on demand its derivative (usually in combination). \qopt{} implements, among other methods, a spectral decomposition or as in this example the default scipy method for calculation of the Fréchet derivative of the matrix exponential. The solver class also interfaces to the filter functions package \cite{filter_func_package}.

\begin{lstlisting}
solver = qo.SchroedingerSMonteCarlo(
    h_ctrl=[.5 * qo.DenseOperator.pauli_x(),
            .5 * qo.DenseOperator.pauli_y()],
    h_drift=[0 * qo.DenseOperator.pauli_x()],
    tau=delta_t * np.ones(n_time_steps),
    exponential_method='Frechet',
    h_noise=[.5 * qo.DenseOperator.pauli_z()],
    noise_trace_generator=noise_gen,
    filter_function_h_n=[
        [qo.DenseOperator.pauli_z().data,
        np.ones(n_time_steps)]
    ]
)
\end{lstlisting}
The target operation is an $X_{\pi/2}$-gate and the cost function for the first simulation evaluates the mean deviation between the simulated propagator and the target gate. We can also choose to neglect the systematic errors to calculate the entanglement fidelity on average between the unperturbed simulation and the simulation including noise. The optimization algorithm is per default the gradient-based L-BFGS-B algorithm implemented in scipy and the bounds restrict the search space. 
\begin{lstlisting}
target = (
        qo.DenseOperator.pauli_x()
    ).exp(.25j * np.pi)

cost_func_qs = qo.OperationNoiseInfidelity(
    solver=solver,
    target=target,
    neglect_systematic_errors=False,
    label=[r'$\mathcal{I}_{\mathrm{qs}}$', ]
)

optimizer_qs = qo.ScalarMinimizingOptimizer(
    system_simulator=qo.Simulator(
        solvers=[solver, ],
        cost_funcs=[cost_func_qs, ]
    ),
    bounds=[[-1, 1], ] * 2 * n_time_steps
)
\end{lstlisting}
In the second simulation we use one cost function for the systematic deviations calculated by the standard entanglement fidelity and a second cost function calculating the infidelity caused by pink noise based on filter functions. The sampling frequencies for the integral in equation \eqref{eq:entanginfidff} are set at the keyword argument \texttt{omega}.
\begin{lstlisting}
cost_func_plain = qo.OperationInfidelity(
    solver=solver,
    target=target,
    label=[r'$\mathcal{I}_{\mathrm{e}}$']
)
def noise_psd(f):
    return 1e-3 / f

total_time = n_time_steps * delta_t
start = np.log10(1 / total_time)
end = np.log10(1 / delta_t)

cost_func_ff = \
    qo.OperatorFilterFunctionInfidelity(
        solver=solver,
        noise_power_spec_density=noise_psd,
        omega=np.logspace(start, end, 200),
        label=[r'$\mathcal{I}_{\mathrm{ff}}$']
    )

optimizer_ff = qo.ScalarMinimizingOptimizer(
    system_simulator=qo.Simulator(
        solvers=[solver, ],
        cost_funcs=[cost_func_plain,
                    cost_func_ff]
    ),
    bounds=[[-1, 1], ] * 2 * n_time_steps
)
\end{lstlisting}
The simulation is executed with the following commands:
\begin{lstlisting}
np.random.seed(0)
result = optimizer_qs.run_optimization(
    initial_control_amplitudes=
    np.random.rand(20, 2))
result_ff = optimizer_ff.run_optimization(
    initial_control_amplitudes=
    result.final_parameters)
\end{lstlisting}
The resulting data can be stored with the data container class and plotted with the analyzer class.
\begin{lstlisting}
data_qs = qo.DataContainer()
data_qs.append_optim_result(result)
analyser_qs = qo.Analyser(data_qs)
data_ff = qo.DataContainer()
data_ff.append_optim_result(result_ff)
analyser_ff = qo.Analyser(data_ff)
\end{lstlisting}
Some cosmetics in plotting commands are shorted for the sake of brevity. We can plot a final pulse (compare to Fig.~\ref{fig:pulse_plots}) with:
\begin{lstlisting}
solver.transfer_function.plot_pulse(result.final_parameters)
\end{lstlisting}
And the infidelity during the optimization (compare to Fig.~\ref{fig:optimization_example}) with:
\begin{lstlisting}
fig, axes = plt.subplots(2)
analyser_qs.plot_costs(ax=axes[0])
analyser_ff.plot_costs(ax=axes[1])
\end{lstlisting}

Each optimization took less than $\SI{10}{\s}$ on a desktop PC.
The decrease in the infidelities during the optimizations can be seen in Fig.\,\ref{fig:optimization_example} and the final pulses are plotted in Fig.\,\ref{fig:pulse_plots}.
With the second optimization run, the infidelity as calculated in Eq.~\eqref{eq:entanginfidff} is decreased by about a factor of six, demonstrating the benefit of explicitly considering fast auto-correlated noise.

\section{Summary and Outlook}
\label{sec:summary and outlook}

Our open-source software package \qopt{} \cite{qopt} provides a general platform for the realistic simulation of qubit systems and QOC. We set the focus on the accurate and efficient simulation of arbitrary classical noise by including three noise simulation methods with distinct application areas. Quasi-static noise is efficiently simulated with Monte Carlo methods or numerical integration, Markovian noise is described best by a master equation in Lindblad form, while fast non-Markovian noise can be treated with filter functions. In the implementation of each method, the noise model can be drive-dependent. The limitations of control hardware are accounted for by the use of transfer functions.
In addition to the example provided in this paper, the online documentation \cite{qopt-docs} contains a complete API documentation and numerous IPython notebooks discussing \qopt{}'s features and an introduction to practical simulations.

We also provide an open-source repository of public simulation and optimal control projects called \texttt{qopt-applications} \cite{qopt-applications}. It can serve as starting point for new simulations and facilitates the transfer of knowledge and new optimal control techniques. 

QOC will continue to play a role in the search for the optimal qubit system for the construction of universal quantum computers. 
The increasing number of qubits in quantum processors leads to new challenges like the mitigation of crosstalk between adjacent qubits, robustness of quantum gates towards qubit inhomogeneities or increased noise levels in quantum processors operated at higher temperatures.
Since the improvements from adopting QOC can be dramatic, any assessment of a quantum computation platform should take the applicable control techniques into account. 
The clean interface between the simulation and the optimization algorithm make \qopt{} ideal for the comparison of various optimization algorithms in the context of QOC. Novel AI-based optimization algorithms are interesting candidates.

While we have found \qopt{} to be very useful in a number of applications, there is certainly much room for extensions. One such a feature could be the application of \qopt{} for spectral noise analysis. This could, for example, be achieved by the introduction of a cost function class measuring the sensitivity of a pulse towards noise of a specific frequency with the help of filter functions.

If performance becomes a bottleneck, \qopt{} could profit from a high performance compilation with numba \cite{numba}, the use of other algorithms for the calculation of matrix exponentials, or highly optimized implementations of performance critical functions in a compiled language.

For even more general modeling and pulse parameterization capabilities, the amplitude and the transfer function classes could be generalized allowing for example the application of the amplitude function before the transfer function.

We published \qopt{} with the vision of establishing a new community standard for qubit simulations. The application of a common simulation platform makes simulations less time-consuming and more reproducible compared to the use of special-purpose simulation code. Reproducibility increases the trust in simulation results and facilitates the transfer of simulation and optimal control techniques between different qubit systems.
We thus encourage users to upload new simulation code to \texttt{qopt-applications} \cite{qopt-applications} to increase their visibility and contribute to the advancement of the state-of-the-art. We encourage the participation in the development and welcome feedback on which new features would be useful.

\acknowledgements{We thank Alexander Pitchford, Eric Giguère, Neill Lambert and Franco Nori for helpful discussions and their advice in the design of \qopt{}. We also thank Alexander Willmes, Christian Gorjaew, Paul Surrey, Frederike Butt and Jiaqi Ai for testing \qopt{} and providing feedback. We acknowledge support from the European Research Council (ERC) under the European Union’s Horizon 2020 research and innovation programme (grant agreement No 679342), Impulse and Networking Fund of the Helmholtz Association.
}


\bibliographystyle{apsrev4-1}
\bibliography{bibliography, online}

\begin{thebibliography}{53}%
\makeatletter
\providecommand \@ifxundefined [1]{%
 \@ifx{#1\undefined}
}%
\providecommand \@ifnum [1]{%
 \ifnum #1\expandafter \@firstoftwo
 \else \expandafter \@secondoftwo
 \fi
}%
\providecommand \@ifx [1]{%
 \ifx #1\expandafter \@firstoftwo
 \else \expandafter \@secondoftwo
 \fi
}%
\providecommand \natexlab [1]{#1}%
\providecommand \enquote  [1]{``#1''}%
\providecommand \bibnamefont  [1]{#1}%
\providecommand \bibfnamefont [1]{#1}%
\providecommand \citenamefont [1]{#1}%
\providecommand \href@noop [0]{\@secondoftwo}%
\providecommand \href [0]{\begingroup \@sanitize@url \@href}%
\providecommand \@href[1]{\@@startlink{#1}\@@href}%
\providecommand \@@href[1]{\endgroup#1\@@endlink}%
\providecommand \@sanitize@url [0]{\catcode `\\12\catcode `\$12\catcode
  `\&12\catcode `\#12\catcode `\^12\catcode `\_12\catcode `\%12\relax}%
\providecommand \@@startlink[1]{}%
\providecommand \@@endlink[0]{}%
\providecommand \url  [0]{\begingroup\@sanitize@url \@url }%
\providecommand \@url [1]{\endgroup\@href {#1}{\urlprefix }}%
\providecommand \urlprefix  [0]{URL }%
\providecommand \Eprint [0]{\href }%
\providecommand \doibase [0]{http://dx.doi.org/}%
\providecommand \selectlanguage [0]{\@gobble}%
\providecommand \bibinfo  [0]{\@secondoftwo}%
\providecommand \bibfield  [0]{\@secondoftwo}%
\providecommand \translation [1]{[#1]}%
\providecommand \BibitemOpen [0]{}%
\providecommand \bibitemStop [0]{}%
\providecommand \bibitemNoStop [0]{.\EOS\space}%
\providecommand \EOS [0]{\spacefactor3000\relax}%
\providecommand \BibitemShut  [1]{\csname bibitem#1\endcsname}%
\let\auto@bib@innerbib\@empty
\bibitem [{\citenamefont {Unruh}(1995)}]{Unruh95}%
  \BibitemOpen
  \bibfield  {author} {\bibinfo {author} {\bibfnamefont {W.~G.}\ \bibnamefont
  {Unruh}},\ }\href {\doibase 10.1103/PhysRevA.51.992} {\bibfield  {journal}
  {\bibinfo  {journal} {Phys. Rev. A}\ }\textbf {\bibinfo {volume} {51}},\
  \bibinfo {pages} {992} (\bibinfo {year} {1995})}\BibitemShut {NoStop}%
\bibitem [{\citenamefont {Glaser}\ \emph {et~al.}(2015)\citenamefont {Glaser},
  \citenamefont {Boscain}, \citenamefont {Calarco}, \citenamefont {Koch},
  \citenamefont {K{\"{o}}ckenberger}, \citenamefont {Kosloff}, \citenamefont
  {Kuprov}, \citenamefont {Luy}, \citenamefont {Schirmer}, \citenamefont
  {Schulte-Herbr{\"{u}}ggen}, \citenamefont {Sugny},\ and\ \citenamefont
  {Wilhelm}}]{TrainSchroedCat}%
  \BibitemOpen
  \bibfield  {author} {\bibinfo {author} {\bibfnamefont {S.~J.}\ \bibnamefont
  {Glaser}}, \bibinfo {author} {\bibfnamefont {U.}~\bibnamefont {Boscain}},
  \bibinfo {author} {\bibfnamefont {T.}~\bibnamefont {Calarco}}, \bibinfo
  {author} {\bibfnamefont {C.~P.}\ \bibnamefont {Koch}}, \bibinfo {author}
  {\bibfnamefont {W.}~\bibnamefont {K{\"{o}}ckenberger}}, \bibinfo {author}
  {\bibfnamefont {R.}~\bibnamefont {Kosloff}}, \bibinfo {author} {\bibfnamefont
  {I.}~\bibnamefont {Kuprov}}, \bibinfo {author} {\bibfnamefont
  {B.}~\bibnamefont {Luy}}, \bibinfo {author} {\bibfnamefont {S.}~\bibnamefont
  {Schirmer}}, \bibinfo {author} {\bibfnamefont {T.}~\bibnamefont
  {Schulte-Herbr{\"{u}}ggen}}, \bibinfo {author} {\bibfnamefont
  {D.}~\bibnamefont {Sugny}}, \ and\ \bibinfo {author} {\bibfnamefont {F.~K.}\
  \bibnamefont {Wilhelm}},\ }\href@noop {} {\bibfield  {journal} {\bibinfo
  {journal} {European Physical Journal D}\ }\textbf {\bibinfo {volume} {69}}
  (\bibinfo {year} {2015})}\BibitemShut {NoStop}%
\bibitem [{\citenamefont {Brif}\ \emph {et~al.}(2010)\citenamefont {Brif},
  \citenamefont {Chakrabarti},\ and\ \citenamefont
  {Rabitz}}]{qoc_outlook_2010}%
  \BibitemOpen
  \bibfield  {author} {\bibinfo {author} {\bibfnamefont {C.}~\bibnamefont
  {Brif}}, \bibinfo {author} {\bibfnamefont {R.}~\bibnamefont {Chakrabarti}}, \
  and\ \bibinfo {author} {\bibfnamefont {H.}~\bibnamefont {Rabitz}},\ }\href
  {\doibase 10.1088/1367-2630/12/7/075008} {\bibfield  {journal} {\bibinfo
  {journal} {New Journal of Physics}\ }\textbf {\bibinfo {volume} {12}},\
  \bibinfo {pages} {075008} (\bibinfo {year} {2010})}\BibitemShut {NoStop}%
\bibitem [{\citenamefont {Chow}\ \emph {et~al.}(2010)\citenamefont {Chow},
  \citenamefont {DiCarlo}, \citenamefont {Gambetta}, \citenamefont {Motzoi},
  \citenamefont {Frunzio}, \citenamefont {Girvin},\ and\ \citenamefont
  {Schoelkopf}}]{optimizedPulsesTransmon}%
  \BibitemOpen
  \bibfield  {author} {\bibinfo {author} {\bibfnamefont {J.~M.}\ \bibnamefont
  {Chow}}, \bibinfo {author} {\bibfnamefont {L.}~\bibnamefont {DiCarlo}},
  \bibinfo {author} {\bibfnamefont {J.~M.}\ \bibnamefont {Gambetta}}, \bibinfo
  {author} {\bibfnamefont {F.}~\bibnamefont {Motzoi}}, \bibinfo {author}
  {\bibfnamefont {L.}~\bibnamefont {Frunzio}}, \bibinfo {author} {\bibfnamefont
  {S.~M.}\ \bibnamefont {Girvin}}, \ and\ \bibinfo {author} {\bibfnamefont
  {R.~J.}\ \bibnamefont {Schoelkopf}},\ }\href {\doibase
  10.1103/PhysRevA.82.040305} {\bibfield  {journal} {\bibinfo  {journal} {Phys.
  Rev. A}\ }\textbf {\bibinfo {volume} {82}},\ \bibinfo {pages} {040305}
  (\bibinfo {year} {2010})}\BibitemShut {NoStop}%
\bibitem [{\citenamefont {Kelly}\ \emph
  {et~al.}(2014{\natexlab{a}})\citenamefont {Kelly}, \citenamefont {Barends},
  \citenamefont {Campbell}, \citenamefont {Chen}, \citenamefont {Chen},
  \citenamefont {Chiaro}, \citenamefont {Dunsworth}, \citenamefont {Fowler},
  \citenamefont {Hoi}, \citenamefont {Jeffrey}, \citenamefont {Megrant},
  \citenamefont {Mutus}, \citenamefont {Neill}, \citenamefont {O'Malley},
  \citenamefont {Quintana}, \citenamefont {Roushan}, \citenamefont {Sank},
  \citenamefont {Vainsencher}, \citenamefont {Wenner}, \citenamefont {White},
  \citenamefont {Cleland},\ and\ \citenamefont
  {Martinis}}]{OQCwithrandomizedbenchmarking}%
  \BibitemOpen
  \bibfield  {author} {\bibinfo {author} {\bibfnamefont {J.}~\bibnamefont
  {Kelly}}, \bibinfo {author} {\bibfnamefont {R.}~\bibnamefont {Barends}},
  \bibinfo {author} {\bibfnamefont {B.}~\bibnamefont {Campbell}}, \bibinfo
  {author} {\bibfnamefont {Y.}~\bibnamefont {Chen}}, \bibinfo {author}
  {\bibfnamefont {Z.}~\bibnamefont {Chen}}, \bibinfo {author} {\bibfnamefont
  {B.}~\bibnamefont {Chiaro}}, \bibinfo {author} {\bibfnamefont
  {A.}~\bibnamefont {Dunsworth}}, \bibinfo {author} {\bibfnamefont {A.~G.}\
  \bibnamefont {Fowler}}, \bibinfo {author} {\bibfnamefont {I.-C.}\
  \bibnamefont {Hoi}}, \bibinfo {author} {\bibfnamefont {E.}~\bibnamefont
  {Jeffrey}}, \bibinfo {author} {\bibfnamefont {A.}~\bibnamefont {Megrant}},
  \bibinfo {author} {\bibfnamefont {J.}~\bibnamefont {Mutus}}, \bibinfo
  {author} {\bibfnamefont {C.}~\bibnamefont {Neill}}, \bibinfo {author}
  {\bibfnamefont {P.~J.~J.}\ \bibnamefont {O'Malley}}, \bibinfo {author}
  {\bibfnamefont {C.}~\bibnamefont {Quintana}}, \bibinfo {author}
  {\bibfnamefont {P.}~\bibnamefont {Roushan}}, \bibinfo {author} {\bibfnamefont
  {D.}~\bibnamefont {Sank}}, \bibinfo {author} {\bibfnamefont {A.}~\bibnamefont
  {Vainsencher}}, \bibinfo {author} {\bibfnamefont {J.}~\bibnamefont {Wenner}},
  \bibinfo {author} {\bibfnamefont {T.~C.}\ \bibnamefont {White}}, \bibinfo
  {author} {\bibfnamefont {A.~N.}\ \bibnamefont {Cleland}}, \ and\ \bibinfo
  {author} {\bibfnamefont {J.~M.}\ \bibnamefont {Martinis}},\ }\href {\doibase
  10.1103/PhysRevLett.112.240504} {\bibfield  {journal} {\bibinfo  {journal}
  {Phys. Rev. Lett.}\ }\textbf {\bibinfo {volume} {112}},\ \bibinfo {pages}
  {240504} (\bibinfo {year} {2014}{\natexlab{a}})}\BibitemShut {NoStop}%
\bibitem [{\citenamefont {Preskill}(2018)}]{Preskill2018}%
  \BibitemOpen
  \bibfield  {author} {\bibinfo {author} {\bibfnamefont {J.}~\bibnamefont
  {Preskill}},\ }\href {\doibase 10.22331/q-2018-08-06-79} {\bibfield
  {journal} {\bibinfo  {journal} {{Quantum}}\ }\textbf {\bibinfo {volume}
  {2}},\ \bibinfo {pages} {79} (\bibinfo {year} {2018})}\BibitemShut {NoStop}%
\bibitem [{\citenamefont {Wang}\ \emph {et~al.}(2014)\citenamefont {Wang},
  \citenamefont {Bishop}, \citenamefont {Barnes}, \citenamefont {Kestner},\
  and\ \citenamefont {Sarma}}]{Wang2014}%
  \BibitemOpen
  \bibfield  {author} {\bibinfo {author} {\bibfnamefont {X.}~\bibnamefont
  {Wang}}, \bibinfo {author} {\bibfnamefont {L.~S.}\ \bibnamefont {Bishop}},
  \bibinfo {author} {\bibfnamefont {E.}~\bibnamefont {Barnes}}, \bibinfo
  {author} {\bibfnamefont {J.~P.}\ \bibnamefont {Kestner}}, \ and\ \bibinfo
  {author} {\bibfnamefont {S.~D.}\ \bibnamefont {Sarma}},\ }\href {\doibase
  10.1103/PhysRevA.89.022310} {\bibfield  {journal} {\bibinfo  {journal} {Phys.
  Rev. A}\ }\textbf {\bibinfo {volume} {89}},\ \bibinfo {pages} {022310}
  (\bibinfo {year} {2014})}\BibitemShut {NoStop}%
\bibitem [{\citenamefont {Yang}\ and\ \citenamefont {Wang}(2016)}]{Yang2016}%
  \BibitemOpen
  \bibfield  {author} {\bibinfo {author} {\bibfnamefont {X.~C.}\ \bibnamefont
  {Yang}}\ and\ \bibinfo {author} {\bibfnamefont {X.}~\bibnamefont {Wang}},\
  }\href {\doibase 10.1038/srep28996} {\bibfield  {journal} {\bibinfo
  {journal} {Sci. Rep.}\ }\textbf {\bibinfo {volume} {6}},\ \bibinfo {pages}
  {1} (\bibinfo {year} {2016})}\BibitemShut {NoStop}%
\bibitem [{\citenamefont {Anderson}\ \emph {et~al.}(2015)\citenamefont
  {Anderson}, \citenamefont {Sosa-Martinez}, \citenamefont {Riofr\'{\i}o},
  \citenamefont {Deutsch},\ and\ \citenamefont {Jessen}}]{Anderson2015}%
  \BibitemOpen
  \bibfield  {author} {\bibinfo {author} {\bibfnamefont {B.~E.}\ \bibnamefont
  {Anderson}}, \bibinfo {author} {\bibfnamefont {H.}~\bibnamefont
  {Sosa-Martinez}}, \bibinfo {author} {\bibfnamefont {C.~A.}\ \bibnamefont
  {Riofr\'{\i}o}}, \bibinfo {author} {\bibfnamefont {I.~H.}\ \bibnamefont
  {Deutsch}}, \ and\ \bibinfo {author} {\bibfnamefont {P.~S.}\ \bibnamefont
  {Jessen}},\ }\href {\doibase 10.1103/PhysRevLett.114.240401} {\bibfield
  {journal} {\bibinfo  {journal} {Phys. Rev. Lett.}\ }\textbf {\bibinfo
  {volume} {114}},\ \bibinfo {pages} {240401} (\bibinfo {year}
  {2015})}\BibitemShut {NoStop}%
\bibitem [{\citenamefont {Floether}\ \emph {et~al.}(2012)\citenamefont
  {Floether}, \citenamefont {de~Fouquieres},\ and\ \citenamefont
  {Schirmer}}]{Floether_2012MarkVsNonMark}%
  \BibitemOpen
  \bibfield  {author} {\bibinfo {author} {\bibfnamefont {F.~F.}\ \bibnamefont
  {Floether}}, \bibinfo {author} {\bibfnamefont {P.}~\bibnamefont
  {de~Fouquieres}}, \ and\ \bibinfo {author} {\bibfnamefont {S.~G.}\
  \bibnamefont {Schirmer}},\ }\href {\doibase 10.1088/1367-2630/14/7/073023}
  {\bibfield  {journal} {\bibinfo  {journal} {New Journal of Physics}\ }\textbf
  {\bibinfo {volume} {14}},\ \bibinfo {pages} {073023} (\bibinfo {year}
  {2012})}\BibitemShut {NoStop}%
\bibitem [{\citenamefont {Geck}\ \emph {et~al.}(2019)\citenamefont {Geck},
  \citenamefont {Kruth}, \citenamefont {Bluhm}, \citenamefont {van Waasen},\
  and\ \citenamefont {Heinen}}]{Geck_2019}%
  \BibitemOpen
  \bibfield  {author} {\bibinfo {author} {\bibfnamefont {L.}~\bibnamefont
  {Geck}}, \bibinfo {author} {\bibfnamefont {A.}~\bibnamefont {Kruth}},
  \bibinfo {author} {\bibfnamefont {H.}~\bibnamefont {Bluhm}}, \bibinfo
  {author} {\bibfnamefont {S.}~\bibnamefont {van Waasen}}, \ and\ \bibinfo
  {author} {\bibfnamefont {S.}~\bibnamefont {Heinen}},\ }\href {\doibase
  10.1088/2058-9565/ab5e07} {\bibfield  {journal} {\bibinfo  {journal} {Quantum
  Science and Technology}\ }\textbf {\bibinfo {volume} {5}},\ \bibinfo {pages}
  {015004} (\bibinfo {year} {2019})}\BibitemShut {NoStop}%
\bibitem [{\citenamefont {{Charbon}}\ \emph {et~al.}(2016)\citenamefont
  {{Charbon}}, \citenamefont {{Sebastiano}}, \citenamefont {{Vladimirescu}},
  \citenamefont {{Homulle}}, \citenamefont {{Visser}}, \citenamefont {{Song}},\
  and\ \citenamefont {{Incandela}}}]{CharbonCryoCMOS}%
  \BibitemOpen
  \bibfield  {author} {\bibinfo {author} {\bibfnamefont {E.}~\bibnamefont
  {{Charbon}}}, \bibinfo {author} {\bibfnamefont {F.}~\bibnamefont
  {{Sebastiano}}}, \bibinfo {author} {\bibfnamefont {A.}~\bibnamefont
  {{Vladimirescu}}}, \bibinfo {author} {\bibfnamefont {H.}~\bibnamefont
  {{Homulle}}}, \bibinfo {author} {\bibfnamefont {S.}~\bibnamefont {{Visser}}},
  \bibinfo {author} {\bibfnamefont {L.}~\bibnamefont {{Song}}}, \ and\ \bibinfo
  {author} {\bibfnamefont {R.~M.}\ \bibnamefont {{Incandela}}},\ }in\ \href
  {\doibase 10.1109/IEDM.2016.7838410} {\emph {\bibinfo {booktitle} {2016 IEEE
  International Electron Devices Meeting (IEDM)}}}\ (\bibinfo {year} {2016})\
  pp.\ \bibinfo {pages} {13.5.1--13.5.4}\BibitemShut {NoStop}%
\bibitem [{\citenamefont {Cerfontaine}\ \emph
  {et~al.}(2020{\natexlab{a}})\citenamefont {Cerfontaine}, \citenamefont
  {Botzem}, \citenamefont {Ritzmann}, \citenamefont {Humpohl}, \citenamefont
  {Ludwig}, \citenamefont {Schuh}, \citenamefont {Bougeard}, \citenamefont
  {Wieck},\ and\ \citenamefont {Bluhm}}]{PascalNatureComms}%
  \BibitemOpen
  \bibfield  {author} {\bibinfo {author} {\bibfnamefont {P.}~\bibnamefont
  {Cerfontaine}}, \bibinfo {author} {\bibfnamefont {T.}~\bibnamefont {Botzem}},
  \bibinfo {author} {\bibfnamefont {J.}~\bibnamefont {Ritzmann}}, \bibinfo
  {author} {\bibfnamefont {S.~S.}\ \bibnamefont {Humpohl}}, \bibinfo {author}
  {\bibfnamefont {A.}~\bibnamefont {Ludwig}}, \bibinfo {author} {\bibfnamefont
  {D.}~\bibnamefont {Schuh}}, \bibinfo {author} {\bibfnamefont
  {D.}~\bibnamefont {Bougeard}}, \bibinfo {author} {\bibfnamefont {A.~D.}\
  \bibnamefont {Wieck}}, \ and\ \bibinfo {author} {\bibfnamefont
  {H.}~\bibnamefont {Bluhm}},\ }\href {\doibase 10.1038/s41467-020-17865-3}
  {\bibfield  {journal} {\bibinfo  {journal} {Nat. Commun.}\ }\textbf {\bibinfo
  {volume} {11}},\ \bibinfo {pages} {4144} (\bibinfo {year}
  {2020}{\natexlab{a}})}\BibitemShut {NoStop}%
\bibitem [{\citenamefont {Cerfontaine}\ \emph
  {et~al.}(2020{\natexlab{b}})\citenamefont {Cerfontaine}, \citenamefont
  {Otten},\ and\ \citenamefont {Bluhm}}]{PascalGSCTheo}%
  \BibitemOpen
  \bibfield  {author} {\bibinfo {author} {\bibfnamefont {P.}~\bibnamefont
  {Cerfontaine}}, \bibinfo {author} {\bibfnamefont {R.}~\bibnamefont {Otten}},
  \ and\ \bibinfo {author} {\bibfnamefont {H.}~\bibnamefont {Bluhm}},\ }\href
  {\doibase 10.1103/PhysRevApplied.13.044071} {\bibfield  {journal} {\bibinfo
  {journal} {Phys. Rev. Applied}\ }\textbf {\bibinfo {volume} {13}},\ \bibinfo
  {pages} {044071} (\bibinfo {year} {2020}{\natexlab{b}})}\BibitemShut
  {NoStop}%
\bibitem [{\citenamefont {Egger}\ and\ \citenamefont
  {Wilhelm}(2014)}]{WilhelmAdHOC2014}%
  \BibitemOpen
  \bibfield  {author} {\bibinfo {author} {\bibfnamefont {D.~J.}\ \bibnamefont
  {Egger}}\ and\ \bibinfo {author} {\bibfnamefont {F.~K.}\ \bibnamefont
  {Wilhelm}},\ }\href {\doibase 10.1103/PhysRevLett.112.240503} {\bibfield
  {journal} {\bibinfo  {journal} {Phys. Rev. Lett.}\ }\textbf {\bibinfo
  {volume} {112}},\ \bibinfo {pages} {240503} (\bibinfo {year}
  {2014})}\BibitemShut {NoStop}%
\bibitem [{\citenamefont {Cerfontaine}\ \emph
  {et~al.}(2020{\natexlab{c}})\citenamefont {Cerfontaine}, \citenamefont
  {Otten}, \citenamefont {Wolfe}, \citenamefont {Bethke},\ and\ \citenamefont
  {Bluhm}}]{PascalHighFidGates}%
  \BibitemOpen
  \bibfield  {author} {\bibinfo {author} {\bibfnamefont {P.}~\bibnamefont
  {Cerfontaine}}, \bibinfo {author} {\bibfnamefont {R.}~\bibnamefont {Otten}},
  \bibinfo {author} {\bibfnamefont {M.~A.}\ \bibnamefont {Wolfe}}, \bibinfo
  {author} {\bibfnamefont {P.}~\bibnamefont {Bethke}}, \ and\ \bibinfo {author}
  {\bibfnamefont {H.}~\bibnamefont {Bluhm}},\ }\href {\doibase
  10.1103/PhysRevB.101.155311} {\bibfield  {journal} {\bibinfo  {journal}
  {Phys. Rev. B}\ }\textbf {\bibinfo {volume} {101}},\ \bibinfo {pages}
  {155311} (\bibinfo {year} {2020}{\natexlab{c}})}\BibitemShut {NoStop}%
\bibitem [{\citenamefont {Cerfontaine}\ \emph {et~al.}(2014)\citenamefont
  {Cerfontaine}, \citenamefont {Botzem}, \citenamefont {DiVincenzo},\ and\
  \citenamefont {Bluhm}}]{PascalHighFidSingleQubitGatesTheo}%
  \BibitemOpen
  \bibfield  {author} {\bibinfo {author} {\bibfnamefont {P.}~\bibnamefont
  {Cerfontaine}}, \bibinfo {author} {\bibfnamefont {T.}~\bibnamefont {Botzem}},
  \bibinfo {author} {\bibfnamefont {D.~P.}\ \bibnamefont {DiVincenzo}}, \ and\
  \bibinfo {author} {\bibfnamefont {H.}~\bibnamefont {Bluhm}},\ }\href
  {\doibase 10.1103/PhysRevLett.113.150501} {\bibfield  {journal} {\bibinfo
  {journal} {Phys. Rev. Lett.}\ }\textbf {\bibinfo {volume} {113}},\ \bibinfo
  {pages} {150501} (\bibinfo {year} {2014})}\BibitemShut {NoStop}%
\bibitem [{\citenamefont {Wittler}\ \emph {et~al.}(2020)\citenamefont
  {Wittler}, \citenamefont {Roy}, \citenamefont {Pack}, \citenamefont
  {Werninghaus}, \citenamefont {Roy}, \citenamefont {Egger}, \citenamefont
  {Filipp}, \citenamefont {Wilhelm},\ and\ \citenamefont
  {Machnes}}]{C3wittler2020integrated}%
  \BibitemOpen
  \bibfield  {author} {\bibinfo {author} {\bibfnamefont {N.}~\bibnamefont
  {Wittler}}, \bibinfo {author} {\bibfnamefont {F.}~\bibnamefont {Roy}},
  \bibinfo {author} {\bibfnamefont {K.}~\bibnamefont {Pack}}, \bibinfo {author}
  {\bibfnamefont {M.}~\bibnamefont {Werninghaus}}, \bibinfo {author}
  {\bibfnamefont {A.~S.}\ \bibnamefont {Roy}}, \bibinfo {author} {\bibfnamefont
  {D.~J.}\ \bibnamefont {Egger}}, \bibinfo {author} {\bibfnamefont
  {S.}~\bibnamefont {Filipp}}, \bibinfo {author} {\bibfnamefont {F.~K.}\
  \bibnamefont {Wilhelm}}, \ and\ \bibinfo {author} {\bibfnamefont
  {S.}~\bibnamefont {Machnes}},\ }\href@noop {} {\enquote {\bibinfo {title} {An
  integrated tool-set for control, calibration and characterization of quantum
  devices applied to superconducting qubits},}\ } (\bibinfo {year} {2020}),\
  \Eprint {http://arxiv.org/abs/2009.09866} {arXiv:2009.09866 [quant-ph]}
  \BibitemShut {NoStop}%
\bibitem [{\citenamefont {Motzoi}\ \emph {et~al.}(2011)\citenamefont {Motzoi},
  \citenamefont {Gambetta}, \citenamefont {Merkel},\ and\ \citenamefont
  {Wilhelm}}]{TransferFunc}%
  \BibitemOpen
  \bibfield  {author} {\bibinfo {author} {\bibfnamefont {F.}~\bibnamefont
  {Motzoi}}, \bibinfo {author} {\bibfnamefont {J.~M.}\ \bibnamefont
  {Gambetta}}, \bibinfo {author} {\bibfnamefont {S.~T.}\ \bibnamefont
  {Merkel}}, \ and\ \bibinfo {author} {\bibfnamefont {F.~K.}\ \bibnamefont
  {Wilhelm}},\ }\href {\doibase 10.1103/PhysRevA.84.022307} {\bibfield
  {journal} {\bibinfo  {journal} {Phys. Rev. A}\ }\textbf {\bibinfo {volume}
  {84}},\ \bibinfo {pages} {022307} (\bibinfo {year} {2011})}\BibitemShut
  {NoStop}%
\bibitem [{\citenamefont {Kelly}\ \emph
  {et~al.}(2014{\natexlab{b}})\citenamefont {Kelly}, \citenamefont {Barends},
  \citenamefont {Campbell}, \citenamefont {Chen}, \citenamefont {Chen},
  \citenamefont {Chiaro}, \citenamefont {Dunsworth}, \citenamefont {Fowler},
  \citenamefont {Hoi}, \citenamefont {Jeffrey}, \citenamefont {Megrant},
  \citenamefont {Mutus}, \citenamefont {Neill}, \citenamefont {O'Malley},
  \citenamefont {Quintana}, \citenamefont {Roushan}, \citenamefont {Sank},
  \citenamefont {Vainsencher}, \citenamefont {Wenner}, \citenamefont {White},
  \citenamefont {Cleland},\ and\ \citenamefont {Martinis}}]{KellyBT2014}%
  \BibitemOpen
  \bibfield  {author} {\bibinfo {author} {\bibfnamefont {J.}~\bibnamefont
  {Kelly}}, \bibinfo {author} {\bibfnamefont {R.}~\bibnamefont {Barends}},
  \bibinfo {author} {\bibfnamefont {B.}~\bibnamefont {Campbell}}, \bibinfo
  {author} {\bibfnamefont {Y.}~\bibnamefont {Chen}}, \bibinfo {author}
  {\bibfnamefont {Z.}~\bibnamefont {Chen}}, \bibinfo {author} {\bibfnamefont
  {B.}~\bibnamefont {Chiaro}}, \bibinfo {author} {\bibfnamefont
  {A.}~\bibnamefont {Dunsworth}}, \bibinfo {author} {\bibfnamefont {A.~G.}\
  \bibnamefont {Fowler}}, \bibinfo {author} {\bibfnamefont {I.-C.}\
  \bibnamefont {Hoi}}, \bibinfo {author} {\bibfnamefont {E.}~\bibnamefont
  {Jeffrey}}, \bibinfo {author} {\bibfnamefont {A.}~\bibnamefont {Megrant}},
  \bibinfo {author} {\bibfnamefont {J.}~\bibnamefont {Mutus}}, \bibinfo
  {author} {\bibfnamefont {C.}~\bibnamefont {Neill}}, \bibinfo {author}
  {\bibfnamefont {P.~J.~J.}\ \bibnamefont {O'Malley}}, \bibinfo {author}
  {\bibfnamefont {C.}~\bibnamefont {Quintana}}, \bibinfo {author}
  {\bibfnamefont {P.}~\bibnamefont {Roushan}}, \bibinfo {author} {\bibfnamefont
  {D.}~\bibnamefont {Sank}}, \bibinfo {author} {\bibfnamefont {A.}~\bibnamefont
  {Vainsencher}}, \bibinfo {author} {\bibfnamefont {J.}~\bibnamefont {Wenner}},
  \bibinfo {author} {\bibfnamefont {T.~C.}\ \bibnamefont {White}}, \bibinfo
  {author} {\bibfnamefont {A.~N.}\ \bibnamefont {Cleland}}, \ and\ \bibinfo
  {author} {\bibfnamefont {J.~M.}\ \bibnamefont {Martinis}},\ }\href {\doibase
  10.1103/PhysRevLett.112.240504} {\bibfield  {journal} {\bibinfo  {journal}
  {Phys. Rev. Lett.}\ }\textbf {\bibinfo {volume} {112}},\ \bibinfo {pages}
  {240504} (\bibinfo {year} {2014}{\natexlab{b}})}\BibitemShut {NoStop}%
\bibitem [{\citenamefont {Dial}\ \emph {et~al.}(2013)\citenamefont {Dial},
  \citenamefont {Shulman}, \citenamefont {Harvey}, \citenamefont {Bluhm},
  \citenamefont {Umansky},\ and\ \citenamefont
  {Yacoby}}]{ChargeNoiseSpectrPhysRevLett.110.146804}%
  \BibitemOpen
  \bibfield  {author} {\bibinfo {author} {\bibfnamefont {O.~E.}\ \bibnamefont
  {Dial}}, \bibinfo {author} {\bibfnamefont {M.~D.}\ \bibnamefont {Shulman}},
  \bibinfo {author} {\bibfnamefont {S.~P.}\ \bibnamefont {Harvey}}, \bibinfo
  {author} {\bibfnamefont {H.}~\bibnamefont {Bluhm}}, \bibinfo {author}
  {\bibfnamefont {V.}~\bibnamefont {Umansky}}, \ and\ \bibinfo {author}
  {\bibfnamefont {A.}~\bibnamefont {Yacoby}},\ }\href {\doibase
  10.1103/PhysRevLett.110.146804} {\bibfield  {journal} {\bibinfo  {journal}
  {Phys. Rev. Lett.}\ }\textbf {\bibinfo {volume} {110}},\ \bibinfo {pages}
  {146804} (\bibinfo {year} {2013})}\BibitemShut {NoStop}%
\bibitem [{\citenamefont {Yoneda}\ \emph {et~al.}(2018)\citenamefont {Yoneda},
  \citenamefont {Takeda}, \citenamefont {Otsuka}, \citenamefont {Nakajima},
  \citenamefont {Delbecq}, \citenamefont {Allison}, \citenamefont {Honda},
  \citenamefont {Kodera}, \citenamefont {Oda}, \citenamefont {Hoshi},
  \citenamefont {Usami}, \citenamefont {Itoh},\ and\ \citenamefont
  {Tarucha}}]{999Yoneda2018}%
  \BibitemOpen
  \bibfield  {author} {\bibinfo {author} {\bibfnamefont {J.}~\bibnamefont
  {Yoneda}}, \bibinfo {author} {\bibfnamefont {K.}~\bibnamefont {Takeda}},
  \bibinfo {author} {\bibfnamefont {T.}~\bibnamefont {Otsuka}}, \bibinfo
  {author} {\bibfnamefont {T.}~\bibnamefont {Nakajima}}, \bibinfo {author}
  {\bibfnamefont {M.~R.}\ \bibnamefont {Delbecq}}, \bibinfo {author}
  {\bibfnamefont {G.}~\bibnamefont {Allison}}, \bibinfo {author} {\bibfnamefont
  {T.}~\bibnamefont {Honda}}, \bibinfo {author} {\bibfnamefont
  {T.}~\bibnamefont {Kodera}}, \bibinfo {author} {\bibfnamefont
  {S.}~\bibnamefont {Oda}}, \bibinfo {author} {\bibfnamefont {Y.}~\bibnamefont
  {Hoshi}}, \bibinfo {author} {\bibfnamefont {N.}~\bibnamefont {Usami}},
  \bibinfo {author} {\bibfnamefont {K.~M.}\ \bibnamefont {Itoh}}, \ and\
  \bibinfo {author} {\bibfnamefont {S.}~\bibnamefont {Tarucha}},\ }\href
  {\doibase 10.1038/s41565-017-0014-x} {\bibfield  {journal} {\bibinfo
  {journal} {Nature Nanotechnology}\ }\textbf {\bibinfo {volume} {13}},\
  \bibinfo {pages} {102} (\bibinfo {year} {2018})}\BibitemShut {NoStop}%
\bibitem [{\citenamefont {Khaneja}\ \emph {et~al.}(2005)\citenamefont
  {Khaneja}, \citenamefont {Reiss}, \citenamefont {Kehlet}, \citenamefont
  {Schulte-Herbrüggen},\ and\ \citenamefont {Glaser}}]{grape}%
  \BibitemOpen
  \bibfield  {author} {\bibinfo {author} {\bibfnamefont {N.}~\bibnamefont
  {Khaneja}}, \bibinfo {author} {\bibfnamefont {T.}~\bibnamefont {Reiss}},
  \bibinfo {author} {\bibfnamefont {C.}~\bibnamefont {Kehlet}}, \bibinfo
  {author} {\bibfnamefont {T.}~\bibnamefont {Schulte-Herbrüggen}}, \ and\
  \bibinfo {author} {\bibfnamefont {S.~J.}\ \bibnamefont {Glaser}},\ }\href
  {\doibase https://doi.org/10.1016/j.jmr.2004.11.004} {\bibfield  {journal}
  {\bibinfo  {journal} {Journal of Magnetic Resonance}\ }\textbf {\bibinfo
  {volume} {172}},\ \bibinfo {pages} {296 } (\bibinfo {year}
  {2005})}\BibitemShut {NoStop}%
\bibitem [{\citenamefont {Schirmer}\ and\ \citenamefont
  {de~Fouquieres}(2011)}]{Krotov_Schirmer_2011}%
  \BibitemOpen
  \bibfield  {author} {\bibinfo {author} {\bibfnamefont {S.~G.}\ \bibnamefont
  {Schirmer}}\ and\ \bibinfo {author} {\bibfnamefont {P.}~\bibnamefont
  {de~Fouquieres}},\ }\href {\doibase 10.1088/1367-2630/13/7/073029} {\bibfield
   {journal} {\bibinfo  {journal} {New Journal of Physics}\ }\textbf {\bibinfo
  {volume} {13}},\ \bibinfo {pages} {073029} (\bibinfo {year}
  {2011})}\BibitemShut {NoStop}%
\bibitem [{\citenamefont {Huang}\ and\ \citenamefont {Goan}(2017)}]{Huang2017}%
  \BibitemOpen
  \bibfield  {author} {\bibinfo {author} {\bibfnamefont {C.-H.}\ \bibnamefont
  {Huang}}\ and\ \bibinfo {author} {\bibfnamefont {H.-S.}\ \bibnamefont
  {Goan}},\ }\href {\doibase 10.1103/PhysRevA.95.062325} {\bibfield  {journal}
  {\bibinfo  {journal} {Phys. Rev. A}\ }\textbf {\bibinfo {volume} {95}},\
  \bibinfo {pages} {062325} (\bibinfo {year} {2017})}\BibitemShut {NoStop}%
\bibitem [{\citenamefont {{de Fouquieres}}\ \emph {et~al.}(2011)\citenamefont
  {{de Fouquieres}}, \citenamefont {Schirmer}, \citenamefont {Glaser},\ and\
  \citenamefont {Kuprov}}]{KuprovSecondOrderGradient}%
  \BibitemOpen
  \bibfield  {author} {\bibinfo {author} {\bibfnamefont {P.}~\bibnamefont {{de
  Fouquieres}}}, \bibinfo {author} {\bibfnamefont {S.}~\bibnamefont
  {Schirmer}}, \bibinfo {author} {\bibfnamefont {S.}~\bibnamefont {Glaser}}, \
  and\ \bibinfo {author} {\bibfnamefont {I.}~\bibnamefont {Kuprov}},\ }\href
  {\doibase https://doi.org/10.1016/j.jmr.2011.07.023} {\bibfield  {journal}
  {\bibinfo  {journal} {Journal of Magnetic Resonance}\ }\textbf {\bibinfo
  {volume} {212}},\ \bibinfo {pages} {412} (\bibinfo {year}
  {2011})}\BibitemShut {NoStop}%
\bibitem [{\citenamefont {Machnes}\ \emph {et~al.}(2018)\citenamefont
  {Machnes}, \citenamefont {Ass\'emat}, \citenamefont {Tannor},\ and\
  \citenamefont {Wilhelm}}]{goat}%
  \BibitemOpen
  \bibfield  {author} {\bibinfo {author} {\bibfnamefont {S.}~\bibnamefont
  {Machnes}}, \bibinfo {author} {\bibfnamefont {E.}~\bibnamefont {Ass\'emat}},
  \bibinfo {author} {\bibfnamefont {D.}~\bibnamefont {Tannor}}, \ and\ \bibinfo
  {author} {\bibfnamefont {F.~K.}\ \bibnamefont {Wilhelm}},\ }\href {\doibase
  10.1103/PhysRevLett.120.150401} {\bibfield  {journal} {\bibinfo  {journal}
  {Phys. Rev. Lett.}\ }\textbf {\bibinfo {volume} {120}},\ \bibinfo {pages}
  {150401} (\bibinfo {year} {2018})}\BibitemShut {NoStop}%
\bibitem [{\citenamefont {Teske}\ \emph {et~al.}(2019)\citenamefont {Teske},
  \citenamefont {Humpohl}, \citenamefont {Otten}, \citenamefont {Bethke},
  \citenamefont {Cerfontaine}, \citenamefont {Dedden}, \citenamefont {Ludwig},
  \citenamefont {Wieck},\ and\ \citenamefont {Bluhm}}]{TeskeKalman}%
  \BibitemOpen
  \bibfield  {author} {\bibinfo {author} {\bibfnamefont {J.~D.}\ \bibnamefont
  {Teske}}, \bibinfo {author} {\bibfnamefont {S.~S.}\ \bibnamefont {Humpohl}},
  \bibinfo {author} {\bibfnamefont {R.}~\bibnamefont {Otten}}, \bibinfo
  {author} {\bibfnamefont {P.}~\bibnamefont {Bethke}}, \bibinfo {author}
  {\bibfnamefont {P.}~\bibnamefont {Cerfontaine}}, \bibinfo {author}
  {\bibfnamefont {J.}~\bibnamefont {Dedden}}, \bibinfo {author} {\bibfnamefont
  {A.}~\bibnamefont {Ludwig}}, \bibinfo {author} {\bibfnamefont {A.~D.}\
  \bibnamefont {Wieck}}, \ and\ \bibinfo {author} {\bibfnamefont
  {H.}~\bibnamefont {Bluhm}},\ }\href {\doibase 10.1063/1.5088412} {\bibfield
  {journal} {\bibinfo  {journal} {Applied Physics Letters}\ }\textbf {\bibinfo
  {volume} {114}},\ \bibinfo {pages} {133102} (\bibinfo {year} {2019})},\
  \Eprint {http://arxiv.org/abs/https://doi.org/10.1063/1.5088412}
  {https://doi.org/10.1063/1.5088412} \BibitemShut {NoStop}%
\bibitem [{\citenamefont {Caneva}\ \emph {et~al.}(2011)\citenamefont {Caneva},
  \citenamefont {Calarco},\ and\ \citenamefont {Montangero}}]{crab}%
  \BibitemOpen
  \bibfield  {author} {\bibinfo {author} {\bibfnamefont {T.}~\bibnamefont
  {Caneva}}, \bibinfo {author} {\bibfnamefont {T.}~\bibnamefont {Calarco}}, \
  and\ \bibinfo {author} {\bibfnamefont {S.}~\bibnamefont {Montangero}},\
  }\href {\doibase 10.1103/PhysRevA.84.022326} {\bibfield  {journal} {\bibinfo
  {journal} {Phys. Rev. A}\ }\textbf {\bibinfo {volume} {84}},\ \bibinfo
  {pages} {022326} (\bibinfo {year} {2011})}\BibitemShut {NoStop}%
\bibitem [{\citenamefont {Heck}\ \emph {et~al.}(2018)\citenamefont {Heck},
  \citenamefont {Vuculescu}, \citenamefont {S{\o}rensen}, \citenamefont
  {Zoller}, \citenamefont {Andreasen}, \citenamefont {Bason}, \citenamefont
  {Ejlertsen}, \citenamefont {El{\'\i}asson}, \citenamefont {Haikka},
  \citenamefont {Laustsen}, \citenamefont {Nielsen}, \citenamefont {Mao},
  \citenamefont {M{\"u}ller}, \citenamefont {Napolitano}, \citenamefont
  {Pedersen}, \citenamefont {Thorsen}, \citenamefont {Bergenholtz},
  \citenamefont {Calarco}, \citenamefont {Montangero},\ and\ \citenamefont
  {Sherson}}]{RedCRABHeck}%
  \BibitemOpen
  \bibfield  {author} {\bibinfo {author} {\bibfnamefont {R.}~\bibnamefont
  {Heck}}, \bibinfo {author} {\bibfnamefont {O.}~\bibnamefont {Vuculescu}},
  \bibinfo {author} {\bibfnamefont {J.~J.}\ \bibnamefont {S{\o}rensen}},
  \bibinfo {author} {\bibfnamefont {J.}~\bibnamefont {Zoller}}, \bibinfo
  {author} {\bibfnamefont {M.~G.}\ \bibnamefont {Andreasen}}, \bibinfo {author}
  {\bibfnamefont {M.~G.}\ \bibnamefont {Bason}}, \bibinfo {author}
  {\bibfnamefont {P.}~\bibnamefont {Ejlertsen}}, \bibinfo {author}
  {\bibfnamefont {O.}~\bibnamefont {El{\'\i}asson}}, \bibinfo {author}
  {\bibfnamefont {P.}~\bibnamefont {Haikka}}, \bibinfo {author} {\bibfnamefont
  {J.~S.}\ \bibnamefont {Laustsen}}, \bibinfo {author} {\bibfnamefont {L.~L.}\
  \bibnamefont {Nielsen}}, \bibinfo {author} {\bibfnamefont {A.}~\bibnamefont
  {Mao}}, \bibinfo {author} {\bibfnamefont {R.}~\bibnamefont {M{\"u}ller}},
  \bibinfo {author} {\bibfnamefont {M.}~\bibnamefont {Napolitano}}, \bibinfo
  {author} {\bibfnamefont {M.~K.}\ \bibnamefont {Pedersen}}, \bibinfo {author}
  {\bibfnamefont {A.~R.}\ \bibnamefont {Thorsen}}, \bibinfo {author}
  {\bibfnamefont {C.}~\bibnamefont {Bergenholtz}}, \bibinfo {author}
  {\bibfnamefont {T.}~\bibnamefont {Calarco}}, \bibinfo {author} {\bibfnamefont
  {S.}~\bibnamefont {Montangero}}, \ and\ \bibinfo {author} {\bibfnamefont
  {J.~F.}\ \bibnamefont {Sherson}},\ }\href {\doibase 10.1073/pnas.1716869115}
  {\bibfield  {journal} {\bibinfo  {journal} {Proceedings of the National
  Academy of Sciences}\ }\textbf {\bibinfo {volume} {115}},\ \bibinfo {pages}
  {E11231} (\bibinfo {year} {2018})},\ \Eprint
  {http://arxiv.org/abs/https://www.pnas.org/content/115/48/E11231.full.pdf}
  {https://www.pnas.org/content/115/48/E11231.full.pdf} \BibitemShut {NoStop}%
\bibitem [{\citenamefont {Machnes}\ \emph {et~al.}(2011)\citenamefont
  {Machnes}, \citenamefont {Sander}, \citenamefont {Glaser}, \citenamefont
  {de~Fouqui\`eres}, \citenamefont {Gruslys}, \citenamefont {Schirmer},\ and\
  \citenamefont {Schulte-Herbr\"uggen}}]{DYNAMO}%
  \BibitemOpen
  \bibfield  {author} {\bibinfo {author} {\bibfnamefont {S.}~\bibnamefont
  {Machnes}}, \bibinfo {author} {\bibfnamefont {U.}~\bibnamefont {Sander}},
  \bibinfo {author} {\bibfnamefont {S.~J.}\ \bibnamefont {Glaser}}, \bibinfo
  {author} {\bibfnamefont {P.}~\bibnamefont {de~Fouqui\`eres}}, \bibinfo
  {author} {\bibfnamefont {A.}~\bibnamefont {Gruslys}}, \bibinfo {author}
  {\bibfnamefont {S.}~\bibnamefont {Schirmer}}, \ and\ \bibinfo {author}
  {\bibfnamefont {T.}~\bibnamefont {Schulte-Herbr\"uggen}},\ }\href {\doibase
  10.1103/PhysRevA.84.022305} {\bibfield  {journal} {\bibinfo  {journal} {Phys.
  Rev. A}\ }\textbf {\bibinfo {volume} {84}},\ \bibinfo {pages} {022305}
  (\bibinfo {year} {2011})}\BibitemShut {NoStop}%
\bibitem [{\citenamefont {Johansson}\ \emph {et~al.}(2013)\citenamefont
  {Johansson}, \citenamefont {Nation},\ and\ \citenamefont {Nori}}]{QUTIP}%
  \BibitemOpen
  \bibfield  {author} {\bibinfo {author} {\bibfnamefont {J.}~\bibnamefont
  {Johansson}}, \bibinfo {author} {\bibfnamefont {P.}~\bibnamefont {Nation}}, \
  and\ \bibinfo {author} {\bibfnamefont {F.}~\bibnamefont {Nori}},\ }\href
  {\doibase https://doi.org/10.1016/j.cpc.2012.11.019} {\bibfield  {journal}
  {\bibinfo  {journal} {Computer Physics Communications}\ }\textbf {\bibinfo
  {volume} {184}},\ \bibinfo {pages} {1234 } (\bibinfo {year}
  {2013})}\BibitemShut {NoStop}%
\bibitem [{\citenamefont {Goerz}\ \emph {et~al.}(2019)\citenamefont {Goerz},
  \citenamefont {Basilewitsch}, \citenamefont {Gago-Encinas}, \citenamefont
  {Krauss}, \citenamefont {Horn}, \citenamefont {Reich},\ and\ \citenamefont
  {Koch}}]{Krotov}%
  \BibitemOpen
  \bibfield  {author} {\bibinfo {author} {\bibfnamefont {M.~H.}\ \bibnamefont
  {Goerz}}, \bibinfo {author} {\bibfnamefont {D.}~\bibnamefont {Basilewitsch}},
  \bibinfo {author} {\bibfnamefont {F.}~\bibnamefont {Gago-Encinas}}, \bibinfo
  {author} {\bibfnamefont {M.~G.}\ \bibnamefont {Krauss}}, \bibinfo {author}
  {\bibfnamefont {K.~P.}\ \bibnamefont {Horn}}, \bibinfo {author}
  {\bibfnamefont {D.~M.}\ \bibnamefont {Reich}}, \ and\ \bibinfo {author}
  {\bibfnamefont {C.~P.}\ \bibnamefont {Koch}},\ }\href {\doibase
  10.21468/SciPostPhys.7.6.080} {\bibfield  {journal} {\bibinfo  {journal}
  {SciPost Phys.}\ }\textbf {\bibinfo {volume} {7}},\ \bibinfo {pages} {80}
  (\bibinfo {year} {2019})}\BibitemShut {NoStop}%
\bibitem [{\citenamefont {Sørensen}\ \emph {et~al.}(2019)\citenamefont
  {Sørensen}, \citenamefont {Jensen}, \citenamefont {Heinzel},\ and\
  \citenamefont {Sherson}}]{SORENSEN2019135QEngine}%
  \BibitemOpen
  \bibfield  {author} {\bibinfo {author} {\bibfnamefont {J.}~\bibnamefont
  {Sørensen}}, \bibinfo {author} {\bibfnamefont {J.}~\bibnamefont {Jensen}},
  \bibinfo {author} {\bibfnamefont {T.}~\bibnamefont {Heinzel}}, \ and\
  \bibinfo {author} {\bibfnamefont {J.}~\bibnamefont {Sherson}},\ }\href
  {\doibase https://doi.org/10.1016/j.cpc.2019.04.020} {\bibfield  {journal}
  {\bibinfo  {journal} {Computer Physics Communications}\ }\textbf {\bibinfo
  {volume} {243}},\ \bibinfo {pages} {135} (\bibinfo {year}
  {2019})}\BibitemShut {NoStop}%
\bibitem [{\citenamefont {Silvério}\ \emph {et~al.}(2021)\citenamefont
  {Silvério}, \citenamefont {Grijalva}, \citenamefont {Dalyac}, \citenamefont
  {Leclerc}, \citenamefont {Karalekas}, \citenamefont {Shammah}, \citenamefont
  {Beji}, \citenamefont {Henry},\ and\ \citenamefont {Henriet}}]{pulser}%
  \BibitemOpen
  \bibfield  {author} {\bibinfo {author} {\bibfnamefont {H.}~\bibnamefont
  {Silvério}}, \bibinfo {author} {\bibfnamefont {S.}~\bibnamefont {Grijalva}},
  \bibinfo {author} {\bibfnamefont {C.}~\bibnamefont {Dalyac}}, \bibinfo
  {author} {\bibfnamefont {L.}~\bibnamefont {Leclerc}}, \bibinfo {author}
  {\bibfnamefont {P.~J.}\ \bibnamefont {Karalekas}}, \bibinfo {author}
  {\bibfnamefont {N.}~\bibnamefont {Shammah}}, \bibinfo {author} {\bibfnamefont
  {M.}~\bibnamefont {Beji}}, \bibinfo {author} {\bibfnamefont {L.-P.}\
  \bibnamefont {Henry}}, \ and\ \bibinfo {author} {\bibfnamefont
  {L.}~\bibnamefont {Henriet}},\ }\href@noop {} {\enquote {\bibinfo {title}
  {Pulser: An open-source package for the design of pulse sequences in
  programmable neutral-atom arrays},}\ } (\bibinfo {year} {2021}),\ \Eprint
  {http://arxiv.org/abs/2104.15044} {arXiv:2104.15044 [quant-ph]} \BibitemShut
  {NoStop}%
\bibitem [{\citenamefont {Koch}(2016)}]{Koch_2016robustopensys}%
  \BibitemOpen
  \bibfield  {author} {\bibinfo {author} {\bibfnamefont {C.~P.}\ \bibnamefont
  {Koch}},\ }\href {\doibase 10.1088/0953-8984/28/21/213001} {\bibfield
  {journal} {\bibinfo  {journal} {Journal of Physics: Condensed Matter}\
  }\textbf {\bibinfo {volume} {28}},\ \bibinfo {pages} {213001} (\bibinfo
  {year} {2016})}\BibitemShut {NoStop}%
\bibitem [{\citenamefont {Pawela}\ and\ \citenamefont
  {Sadowski}(2016)}]{Pawela2016methodsdecoherence}%
  \BibitemOpen
  \bibfield  {author} {\bibinfo {author} {\bibfnamefont {L.}~\bibnamefont
  {Pawela}}\ and\ \bibinfo {author} {\bibfnamefont {P.}~\bibnamefont
  {Sadowski}},\ }\href {\doibase 10.1007/s11128-016-1242-y} {\bibfield
  {journal} {\bibinfo  {journal} {Quantum Information Processing}\ }\textbf
  {\bibinfo {volume} {15}},\ \bibinfo {pages} {1937} (\bibinfo {year}
  {2016})}\BibitemShut {NoStop}%
\bibitem [{\citenamefont {Teske}(2020{\natexlab{a}})}]{qopt}%
  \BibitemOpen
  \bibfield  {author} {\bibinfo {author} {\bibfnamefont {J.}~\bibnamefont
  {Teske}},\ }\href@noop {} {\enquote {\bibinfo {title} {{qopt}: A simulation
  and quantum optimal control package},}\ }\bibinfo {howpublished}
  {https://https://github.com/qutech/qopt} (\bibinfo {year}
  {2020}{\natexlab{a}})\BibitemShut {NoStop}%
\bibitem [{\citenamefont {Ball}\ \emph {et~al.}(2020)\citenamefont {Ball},
  \citenamefont {Biercuk}, \citenamefont {Carvalho}, \citenamefont {Chen},
  \citenamefont {Hush}, \citenamefont {Castro}, \citenamefont {Li},
  \citenamefont {Liebermann}, \citenamefont {Slatyer}, \citenamefont {Edmunds},
  \citenamefont {Frey}, \citenamefont {Hempel},\ and\ \citenamefont
  {Milne}}]{QCTRLball2020software}%
  \BibitemOpen
  \bibfield  {author} {\bibinfo {author} {\bibfnamefont {H.}~\bibnamefont
  {Ball}}, \bibinfo {author} {\bibfnamefont {M.~J.}\ \bibnamefont {Biercuk}},
  \bibinfo {author} {\bibfnamefont {A.}~\bibnamefont {Carvalho}}, \bibinfo
  {author} {\bibfnamefont {J.}~\bibnamefont {Chen}}, \bibinfo {author}
  {\bibfnamefont {M.}~\bibnamefont {Hush}}, \bibinfo {author} {\bibfnamefont
  {L.~A.~D.}\ \bibnamefont {Castro}}, \bibinfo {author} {\bibfnamefont
  {L.}~\bibnamefont {Li}}, \bibinfo {author} {\bibfnamefont {P.~J.}\
  \bibnamefont {Liebermann}}, \bibinfo {author} {\bibfnamefont {H.~J.}\
  \bibnamefont {Slatyer}}, \bibinfo {author} {\bibfnamefont {C.}~\bibnamefont
  {Edmunds}}, \bibinfo {author} {\bibfnamefont {V.}~\bibnamefont {Frey}},
  \bibinfo {author} {\bibfnamefont {C.}~\bibnamefont {Hempel}}, \ and\ \bibinfo
  {author} {\bibfnamefont {A.}~\bibnamefont {Milne}},\ }\href@noop {} {\enquote
  {\bibinfo {title} {Software tools for quantum control: Improving quantum
  computer performance through noise and error suppression},}\ } (\bibinfo
  {year} {2020}),\ \Eprint {http://arxiv.org/abs/2001.04060} {arXiv:2001.04060
  [quant-ph]} \BibitemShut {NoStop}%
\bibitem [{\citenamefont {Teske}(2020{\natexlab{b}})}]{qopt-docs}%
  \BibitemOpen
  \bibfield  {author} {\bibinfo {author} {\bibfnamefont {J.}~\bibnamefont
  {Teske}},\ }\href@noop {} {\enquote {\bibinfo {title} {{qopt}: Api
  documentation and introduction notebooks},}\ }\bibinfo {howpublished}
  {https://qopt.readthedocs.io/en/latest/index.html} (\bibinfo {year}
  {2020}{\natexlab{b}})\BibitemShut {NoStop}%
\bibitem [{\citenamefont {Caflish}(1998)}]{CaflischMonteCarloGeneral}%
  \BibitemOpen
  \bibfield  {author} {\bibinfo {author} {\bibfnamefont {R.}~\bibnamefont
  {Caflish}},\ }\href@noop {} {\bibfield  {journal} {\bibinfo  {journal} {Acta
  Numerica}\ ,\ \bibinfo {pages} {1}} (\bibinfo {year} {1998})}\BibitemShut
  {NoStop}%
\bibitem [{\citenamefont {Havel}(2003)}]{ConvertLindblad}%
  \BibitemOpen
  \bibfield  {author} {\bibinfo {author} {\bibfnamefont {T.~F.}\ \bibnamefont
  {Havel}},\ }\href {\doibase 10.1063/1.1518555} {\bibfield  {journal}
  {\bibinfo  {journal} {Journal of Mathematical Physics}\ }\textbf {\bibinfo
  {volume} {44}},\ \bibinfo {pages} {534} (\bibinfo {year} {2003})},\ \Eprint
  {http://arxiv.org/abs/https://aip.scitation.org/doi/pdf/10.1063/1.1518555}
  {https://aip.scitation.org/doi/pdf/10.1063/1.1518555} \BibitemShut {NoStop}%
\bibitem [{\citenamefont {Green}\ \emph {et~al.}(2013)\citenamefont {Green},
  \citenamefont {Sastrawan}, \citenamefont {Uys},\ and\ \citenamefont
  {Biercuk}}]{Green_2013}%
  \BibitemOpen
  \bibfield  {author} {\bibinfo {author} {\bibfnamefont {T.~J.}\ \bibnamefont
  {Green}}, \bibinfo {author} {\bibfnamefont {J.}~\bibnamefont {Sastrawan}},
  \bibinfo {author} {\bibfnamefont {H.}~\bibnamefont {Uys}}, \ and\ \bibinfo
  {author} {\bibfnamefont {M.~J.}\ \bibnamefont {Biercuk}},\ }\href {\doibase
  10.1088/1367-2630/15/9/095004} {\bibfield  {journal} {\bibinfo  {journal}
  {New Journal of Physics}\ }\textbf {\bibinfo {volume} {15}},\ \bibinfo
  {pages} {095004} (\bibinfo {year} {2013})}\BibitemShut {NoStop}%
\bibitem [{\citenamefont {Huang}\ \emph {et~al.}(2019)\citenamefont {Huang},
  \citenamefont {Yang}, \citenamefont {Chen}, \citenamefont {Dzurak},\ and\
  \citenamefont {Goan}}]{ffoptPhysRevA.99.042310}%
  \BibitemOpen
  \bibfield  {author} {\bibinfo {author} {\bibfnamefont {C.-H.}\ \bibnamefont
  {Huang}}, \bibinfo {author} {\bibfnamefont {C.-H.}\ \bibnamefont {Yang}},
  \bibinfo {author} {\bibfnamefont {C.-C.}\ \bibnamefont {Chen}}, \bibinfo
  {author} {\bibfnamefont {A.~S.}\ \bibnamefont {Dzurak}}, \ and\ \bibinfo
  {author} {\bibfnamefont {H.-S.}\ \bibnamefont {Goan}},\ }\href {\doibase
  10.1103/PhysRevA.99.042310} {\bibfield  {journal} {\bibinfo  {journal} {Phys.
  Rev. A}\ }\textbf {\bibinfo {volume} {99}},\ \bibinfo {pages} {042310}
  (\bibinfo {year} {2019})}\BibitemShut {NoStop}%
\bibitem [{\citenamefont {Hangleiter}\ \emph {et~al.}(2021)\citenamefont
  {Hangleiter}, \citenamefont {Cerfontaine},\ and\ \citenamefont
  {Bluhm}}]{hangleiter2021filter}%
  \BibitemOpen
  \bibfield  {author} {\bibinfo {author} {\bibfnamefont {T.}~\bibnamefont
  {Hangleiter}}, \bibinfo {author} {\bibfnamefont {P.}~\bibnamefont
  {Cerfontaine}}, \ and\ \bibinfo {author} {\bibfnamefont {H.}~\bibnamefont
  {Bluhm}},\ }\href@noop {} {\enquote {\bibinfo {title} {Filter function
  formalism and software package to compute quantum processes of gate sequences
  for classical non-markovian noise},}\ } (\bibinfo {year} {2021}),\ \Eprint
  {http://arxiv.org/abs/2103.02403} {arXiv:2103.02403 [quant-ph]} \BibitemShut
  {NoStop}%
\bibitem [{\citenamefont {Cerfontaine}\ \emph {et~al.}(2021)\citenamefont
  {Cerfontaine}, \citenamefont {Hangleiter},\ and\ \citenamefont
  {Bluhm}}]{cerfontaine2021filter}%
  \BibitemOpen
  \bibfield  {author} {\bibinfo {author} {\bibfnamefont {P.}~\bibnamefont
  {Cerfontaine}}, \bibinfo {author} {\bibfnamefont {T.}~\bibnamefont
  {Hangleiter}}, \ and\ \bibinfo {author} {\bibfnamefont {H.}~\bibnamefont
  {Bluhm}},\ }\href@noop {} {\enquote {\bibinfo {title} {Filter functions for
  quantum processes under correlated noise},}\ } (\bibinfo {year} {2021}),\
  \Eprint {http://arxiv.org/abs/2103.02385} {arXiv:2103.02385 [quant-ph]}
  \BibitemShut {NoStop}%
\bibitem [{\citenamefont {Hangleiter}(2019)}]{filter_func_package}%
  \BibitemOpen
  \bibfield  {author} {\bibinfo {author} {\bibfnamefont {T.}~\bibnamefont
  {Hangleiter}},\ }\href@noop {} {\enquote {\bibinfo {title}
  {{filter\_functions}: A package for efficient numerical calculation of
  generalized filter functions},}\ }\bibinfo {howpublished}
  {https://github.com/qutech/filter\_functions} (\bibinfo {year}
  {2019})\BibitemShut {NoStop}%
\bibitem [{\citenamefont {Grace}\ \emph {et~al.}(2010)\citenamefont {Grace},
  \citenamefont {Dominy}, \citenamefont {Kosut}, \citenamefont {Brif},\ and\
  \citenamefont {Rabitz}}]{Grace_2010openfidelity}%
  \BibitemOpen
  \bibfield  {author} {\bibinfo {author} {\bibfnamefont {M.~D.}\ \bibnamefont
  {Grace}}, \bibinfo {author} {\bibfnamefont {J.}~\bibnamefont {Dominy}},
  \bibinfo {author} {\bibfnamefont {R.~L.}\ \bibnamefont {Kosut}}, \bibinfo
  {author} {\bibfnamefont {C.}~\bibnamefont {Brif}}, \ and\ \bibinfo {author}
  {\bibfnamefont {H.}~\bibnamefont {Rabitz}},\ }\href {\doibase
  10.1088/1367-2630/12/1/015001} {\bibfield  {journal} {\bibinfo  {journal}
  {New Journal of Physics}\ }\textbf {\bibinfo {volume} {12}},\ \bibinfo
  {pages} {015001} (\bibinfo {year} {2010})}\BibitemShut {NoStop}%
\bibitem [{\citenamefont {Abdelhafez}\ \emph {et~al.}(2019)\citenamefont
  {Abdelhafez}, \citenamefont {Schuster},\ and\ \citenamefont
  {Koch}}]{TrajectoriesAutodiff}%
  \BibitemOpen
  \bibfield  {author} {\bibinfo {author} {\bibfnamefont {M.}~\bibnamefont
  {Abdelhafez}}, \bibinfo {author} {\bibfnamefont {D.~I.}\ \bibnamefont
  {Schuster}}, \ and\ \bibinfo {author} {\bibfnamefont {J.}~\bibnamefont
  {Koch}},\ }\href {\doibase 10.1103/PhysRevA.99.052327} {\bibfield  {journal}
  {\bibinfo  {journal} {Phys. Rev. A}\ }\textbf {\bibinfo {volume} {99}},\
  \bibinfo {pages} {052327} (\bibinfo {year} {2019})}\BibitemShut {NoStop}%
\bibitem [{\citenamefont {Virtanen}\ \emph {et~al.}(2020)\citenamefont
  {Virtanen}, \citenamefont {Gommers}, \citenamefont {Oliphant}, \citenamefont
  {Haberland}, \citenamefont {Reddy}, \citenamefont {Cournapeau}, \citenamefont
  {Burovski}, \citenamefont {Peterson}, \citenamefont {Weckesser},
  \citenamefont {Bright}, \citenamefont {{van der Walt}}, \citenamefont
  {Brett}, \citenamefont {Wilson}, \citenamefont {Millman}, \citenamefont
  {Mayorov}, \citenamefont {Nelson}, \citenamefont {Jones}, \citenamefont
  {Kern}, \citenamefont {Larson}, \citenamefont {Carey}, \citenamefont {Polat},
  \citenamefont {Feng}, \citenamefont {Moore}, \citenamefont {{VanderPlas}},
  \citenamefont {Laxalde}, \citenamefont {Perktold}, \citenamefont {Cimrman},
  \citenamefont {Henriksen}, \citenamefont {Quintero}, \citenamefont {Harris},
  \citenamefont {Archibald}, \citenamefont {Ribeiro}, \citenamefont
  {Pedregosa}, \citenamefont {{van Mulbregt}},\ and\ \citenamefont {{SciPy 1.0
  Contributors}}}]{scipy}%
  \BibitemOpen
  \bibfield  {author} {\bibinfo {author} {\bibfnamefont {P.}~\bibnamefont
  {Virtanen}}, \bibinfo {author} {\bibfnamefont {R.}~\bibnamefont {Gommers}},
  \bibinfo {author} {\bibfnamefont {T.~E.}\ \bibnamefont {Oliphant}}, \bibinfo
  {author} {\bibfnamefont {M.}~\bibnamefont {Haberland}}, \bibinfo {author}
  {\bibfnamefont {T.}~\bibnamefont {Reddy}}, \bibinfo {author} {\bibfnamefont
  {D.}~\bibnamefont {Cournapeau}}, \bibinfo {author} {\bibfnamefont
  {E.}~\bibnamefont {Burovski}}, \bibinfo {author} {\bibfnamefont
  {P.}~\bibnamefont {Peterson}}, \bibinfo {author} {\bibfnamefont
  {W.}~\bibnamefont {Weckesser}}, \bibinfo {author} {\bibfnamefont
  {J.}~\bibnamefont {Bright}}, \bibinfo {author} {\bibfnamefont {S.~J.}\
  \bibnamefont {{van der Walt}}}, \bibinfo {author} {\bibfnamefont
  {M.}~\bibnamefont {Brett}}, \bibinfo {author} {\bibfnamefont
  {J.}~\bibnamefont {Wilson}}, \bibinfo {author} {\bibfnamefont {K.~J.}\
  \bibnamefont {Millman}}, \bibinfo {author} {\bibfnamefont {N.}~\bibnamefont
  {Mayorov}}, \bibinfo {author} {\bibfnamefont {A.~R.~J.}\ \bibnamefont
  {Nelson}}, \bibinfo {author} {\bibfnamefont {E.}~\bibnamefont {Jones}},
  \bibinfo {author} {\bibfnamefont {R.}~\bibnamefont {Kern}}, \bibinfo {author}
  {\bibfnamefont {E.}~\bibnamefont {Larson}}, \bibinfo {author} {\bibfnamefont
  {C.~J.}\ \bibnamefont {Carey}}, \bibinfo {author} {\bibfnamefont
  {{\.I}.}~\bibnamefont {Polat}}, \bibinfo {author} {\bibfnamefont
  {Y.}~\bibnamefont {Feng}}, \bibinfo {author} {\bibfnamefont {E.~W.}\
  \bibnamefont {Moore}}, \bibinfo {author} {\bibfnamefont {J.}~\bibnamefont
  {{VanderPlas}}}, \bibinfo {author} {\bibfnamefont {D.}~\bibnamefont
  {Laxalde}}, \bibinfo {author} {\bibfnamefont {J.}~\bibnamefont {Perktold}},
  \bibinfo {author} {\bibfnamefont {R.}~\bibnamefont {Cimrman}}, \bibinfo
  {author} {\bibfnamefont {I.}~\bibnamefont {Henriksen}}, \bibinfo {author}
  {\bibfnamefont {E.~A.}\ \bibnamefont {Quintero}}, \bibinfo {author}
  {\bibfnamefont {C.~R.}\ \bibnamefont {Harris}}, \bibinfo {author}
  {\bibfnamefont {A.~M.}\ \bibnamefont {Archibald}}, \bibinfo {author}
  {\bibfnamefont {A.~H.}\ \bibnamefont {Ribeiro}}, \bibinfo {author}
  {\bibfnamefont {F.}~\bibnamefont {Pedregosa}}, \bibinfo {author}
  {\bibfnamefont {P.}~\bibnamefont {{van Mulbregt}}}, \ and\ \bibinfo {author}
  {\bibnamefont {{SciPy 1.0 Contributors}}},\ }\href {\doibase
  10.1038/s41592-019-0686-2} {\bibfield  {journal} {\bibinfo  {journal} {Nature
  Methods}\ }\textbf {\bibinfo {volume} {17}},\ \bibinfo {pages} {261}
  (\bibinfo {year} {2020})}\BibitemShut {NoStop}%
\bibitem [{\citenamefont {Moler}\ and\ \citenamefont
  {Loan}(1978)}]{19matrixexponentials}%
  \BibitemOpen
  \bibfield  {author} {\bibinfo {author} {\bibfnamefont {C.}~\bibnamefont
  {Moler}}\ and\ \bibinfo {author} {\bibfnamefont {C.~V.}\ \bibnamefont
  {Loan}},\ }\href {http://www.jstor.org/stable/2029743} {\bibfield  {journal}
  {\bibinfo  {journal} {SIAM Review}\ }\textbf {\bibinfo {volume} {20}},\
  \bibinfo {pages} {801} (\bibinfo {year} {1978})}\BibitemShut {NoStop}%
\bibitem [{\citenamefont {Teske}(2020{\natexlab{c}})}]{qopt-applications}%
  \BibitemOpen
  \bibfield  {author} {\bibinfo {author} {\bibfnamefont {J.}~\bibnamefont
  {Teske}},\ }\href@noop {} {\enquote {\bibinfo {title} {{qopt-applications}:
  Simulations and optimal control implemented with qopt},}\ }\bibinfo
  {howpublished} {https://github.com/qutech/qopt-applications} (\bibinfo {year}
  {2020}{\natexlab{c}})\BibitemShut {NoStop}%
\bibitem [{\citenamefont {Lam}\ \emph {et~al.}(2015)\citenamefont {Lam},
  \citenamefont {Pitrou},\ and\ \citenamefont {Seibert}}]{numba}%
  \BibitemOpen
  \bibfield  {author} {\bibinfo {author} {\bibfnamefont {S.~K.}\ \bibnamefont
  {Lam}}, \bibinfo {author} {\bibfnamefont {A.}~\bibnamefont {Pitrou}}, \ and\
  \bibinfo {author} {\bibfnamefont {S.}~\bibnamefont {Seibert}},\ }in\ \href
  {\doibase 10.1145/2833157.2833162} {\emph {\bibinfo {booktitle} {Proceedings
  of the Second Workshop on the LLVM Compiler Infrastructure in HPC}}},\
  \bibinfo {series and number} {LLVM '15}\ (\bibinfo  {publisher} {Association
  for Computing Machinery},\ \bibinfo {address} {New York, NY, USA},\ \bibinfo
  {year} {2015})\BibitemShut {NoStop}%
\end{thebibliography}%

\end{document}